\documentstyle[prd,aps,preprint]{revtex}
\tightenlines
\input epsf.tex
\def\DESepsf(#1 width #2){\epsfxsize=#2 \epsfbox{#1}}

\begin{document}

\draft
\preprint{\vbox{
\hbox{UMD-PP-03-22} }}
\title{ ICTP Lectures on Theoretical Aspects of Neutrino Masses and
Mixings\footnote{Lectures delivered at the NEUPAST school at the Abdus
Salam ICTP, Trieste in October, 2002; parts of the material were also
presented at Beyond02 workshop in Oulu, Finland and SUSY02 conference in
DESY, Germany}} 
\author{ R. N. Mohapatra\footnote{e-mail:rmohapat@physics.umd.edu}}

\address{ Department of
Physics, University of Maryland, College Park, MD, 20742, USA}
\date{September, 2002}
\maketitle
\begin{abstract}
Recent neutrino oscillation experiments are yielding valuable information
on the nature of neutrino masses and mixings even though we are far from
a complete understanding of the new physics implied by them. In these
lectures, I summarize the current theoretical status of neutrino mass
physics.
\end{abstract}

\vskip0.5in

\newpage

\noindent {\Large Table of contents}

\bigskip

\noindent {\bf Lecture I}

\begin{quote}
I.1   Introduction\\

I.2   Dirac vrs Majorana masses in two component notation\\

I.3   Experimental evidence for neutrino mass  \\

I.4   Patterns and textures for neutrinos\\

I.5   Solar and atmospheric data and neutrino mass patterns\\

I.6   Mass textures suggested by data\\

I.7   Inverted mass pattern and $L_e-L_\mu-L_\tau$ symmetry\\

I.8   CP violation\\

I.9   Theoretical issues raised by the present neutrino data
\end{quote}

\noindent {\bf Lecture II}

\begin{quote}

II.1  Why neutrino mass implies that we must go beyond the standard model
?\\

II.2  Scenarios for small neutrino masses without right handed neutrinos\\

II.3  Seesaw mechanism and left-right symmetric models for small neutrino
     masses\\

II.4  Gauge model for seesaw mechanism: left-right symmetric unification\\

II.5  Neutrino mixings using seesaw mechanism\\

II.6  General consequences for the seesaw formula for neutrino masses\\

II.7  $L_e-L_\mu-L_\tau$ symmetry and $3\times 2$ seesaw\\

II.8  SO(10) realization of the seesaw mechanism\\

II.9  Naturalness of degenerate neutrinos and type II seesaw formula

\end{quote}

\noindent {\bf Lecture III}

\begin{quote}

III.1  Lepton Flavor violation and neutrino mass\\

III.2  Renormalization group evolution of the neutrino mass matrix\\

III.3  Radiative magnification of neutrino mixings as a way to understand
       the large mixing angles\\

III.4  Example of a neutrino mass matrix unstable under RGE effects\\

III.5  Generating solar mass difference square fron RGE's\\
 
III.6  A horizontal symmetry model for near bimaximal mixing

\end{quote}

\noindent {\bf Lecture IV}

\begin{quote}

IV.1  Bulk neutrinos and neutrino mass in extra dimensional models\\

IV.2  Phenomenological constraints on the bulk neutrino\\

IV.3  Neutrino mass, low fundamental scale and extra dimensions

\end{quote}

\noindent Conclusions and outlook

\newpage
\section{}

\begin{center}
{\bf I.1: Introduction}
\end{center}

For a long time, it was believed that neutrinos are massless, spin half
particles, making them drastically different from their other standard
model spin half cousins such as the charged leptons ($e, \mu,
\tau$) and the quarks ($u,d,s,c,t,b$), which are known to have mass. In
fact the masslessness of the neutrino was considered so sacred in
the 1950s and 1960s that the fundamental law of weak interaction
physics, the successful V-A
theory for charged current weak processes proposed by Sudarshan,
Marshak, Feynman and Gell-Mann was considered to be intimately linked
to this fact. The argument went as such: a massless
fermion field equation is invariant under the $\gamma_5$
transformation; since neutrino is one such particle and it participates
exclusively in weak interactions, the weak interactions must somehow
reflect invariance under the $\gamma_5$-transformation of all fermionic
matter (i.e. quarks, charged
leptons and the neutrinos)  particparting in weak interactions.The
argument is obviously very heuristic; but it is 
not hard to see its profound implication: it leads to the resulting
four-Fermi weak interaction to involve only V-A currents.
This argument remained persuasive  for a long time since there was no
evidence for neutrino mass for almost next 50 years and became a
celebrated myth in particle theory.

This myth has however been shattered by the accumulating evidence for
neutrino mass from the solar and atmospheric neutrino data compiled in the
nineties and still ongoing. One must therefore now be free to look beyond
the $\gamma_5$ invariance idea for exploring new physics as we proceed to
understand the neutrino mass. 

 The possibility of a nonzero neutrino mass at phenomenological level goes
back almost 50 years\cite{history}. In the context of gauge theories, they
were discussed extensively in th 70's and 80's long before there was any
firm evidence for it. For instance the left-right symmetric theories of
weak interactions introduced in 1974 and discussed in those days in
connection with the structure of neutral current weak interactions,
predicted nonzero neutrino mass as a necessary consequence of parity
invariance and quark lepton symmetry.

 The existence of a nonzero neutrino mass makes neutrinos more 
like the quarks, and allows for mixing between the different neutrino
species leading to the phenomenon of neutrino oscillation, an idea
first discussed by Pontecorvo\cite{pont} and Maki et al.\cite{pont} in the
1960's, unleasing a whole
new realm of particle physics phenomena to explore. More importantly,
the simple fact that neutrino masses vanish in
the standard model implies that, evidence for neutrino mass is a solid
evidence for the existence of new physics beyond the standard model. 

We are of course far from a complete picture of the masses and mixings of
the various neutrinos and cannot therefore have a full outline of the
theory
of neutrino masses at present. However there exist enough information and
indirect
indications that constrain the masses and mixings among the neutrinos that
we can see a narrowing of the possibilities for the theories beyond
the standard model. Combined with other ideas outside the neutrino arena
such as supersymmetry and unification, the possibility narrows even
further. Many
clever experiments now under way will soon clarify or rule out many of the
allowed models. It will be one of the goals of this article to give a
panoramic view of the most likely scenarios for new physics that explain 
what is now known about neutrino masses\cite{book}. Such discussions are
of course by nature very subjective and therefore a sincere apology is due
at the beginning of this discussion to all those whose ideas are not cited
in this lecture.

 We hope to emphasize two kinds
of ideas: one which provides a general framework for understanding of the
small neutrino masses, the seesaw mechanism and predicts the
existence of superheavy right handed neutrinos. In the opinion of this
author, these ideas are likely
to be part of the final theory of neutrino masses. We then touch briefly
on some specific models that are based on the above general framework
but attempt to provide an understanding of the detailed mass and mixing
patterns. These
works are instructive for several reasons: first they provide proof of
the detailed workability of the general ideas described above (sort of
existence proofs that things will work); second they often illustrate
the kind of assumptions needed and thru that provide a unique insight into
which
directions the next step should be; finally of course nature may be 
generous in picking one of those models as the final message bearer.

In discussing the neutrino mass, it is instructive to compare it with
the other well known fermion, the electron.
The electron and the neutrino are in many ways very similar particles:
they are both spin half objects; they both participate in weak
interactions with same
strength; in fact they are so similar that in the limit of exact gauge
symmetries they are two states of the same object and therefore in
principle indistinguishable. Yet there are profound differences between
them in the standard model: after gauge symmetry breaking, only
electron has electric charge but the neutrinos come out elctrically
neutral as they should to match observations. Another difference is that
only the lefthanded neutrinos of each generation are included in the
standard
model and not its righthanded counterpart whereas for the $e,\mu,\tau$ and
all the quark flavors,
both helicity states are included. The fact that the righthanded
neutrino is excluded from the standard model coupled with the fact that
$B-L$ is an exact symmetry of the model implies that neutrino remains
massless to all orders in perturbation theory as well as
nonperturbatively, as we will discuss later on in this review.

The fact that the neutrino has no electric charge endows it with certain
properties not shared by other fermions of the standard model. 
One can write two kinds of Lorentz invariant mass terms
for the neutrino, the Dirac and Majorana masses, whereas for the charged
fermions, conservation of electric charge allows only Dirac type mass
terms. In the four component notation for the fermions, the Dirac
mass has the form $\bar{\psi}\psi$, whereas the Majorana mass is of the
form $\psi^TC^{-1}\psi$, where $\psi$ is the four component spinor and
$C$ is the charge conjugation matrix. One can also discuss the two
different kinds of mass terms using the two component notation for the
spinors, which provides a very useful way to discuss neutrino masses.
We therefore present some of the salient concepts behind the two component
description of the neutrino.

\begin{center}
{\bf I.2   Two component notation for neutrinos}
\end{center}
Before we start the discussion of the 2-component neutrino,
let us write down the Dirac equation for an electron\cite{kayser}:
\begin{eqnarray}
i\gamma^{\lambda}\partial_{\lambda}\psi - m\psi =0
\end{eqnarray}
This equation follows from a free Lagrangian
\begin{eqnarray} {\cal L} =
i\bar{\psi}\gamma^{\lambda}\partial_{\lambda}\psi
-m\bar{\psi}\psi
\end{eqnarray}
and leads to the relativistic energy momentum relation
$p^{\lambda}p_{\lambda}
= m^2$ for the spin-half particle only if the four $\gamma_{\lambda}$'s
anticommute. If we take $\gamma_{\lambda}$'s to be $n\times  n$ matrices ,
the
smallest value of $n$ for which four anticommuting matrices exist is four.
Therefore $\psi$ must be a four component spinor. The physical meaning of
the four components is as follows: two components for particle spin up and
down and same for the antiparticle.

A spin-half particle is said to be a Majorana particle if the spinor field
$\psi$ satisfies the condition of being self charge conjugate, i.e.
\begin{eqnarray}
 \psi = \psi^c \equiv C \bar{\psi}^T,
\end{eqnarray}
where $C$ is the charge conjugation matrix and has the property
$C\gamma_{\lambda} C^{-1} = - {\gamma^{\lambda}}^T $.
This constraint reduces the number of independent components of the spinor
by a factor of two, since the particle and the antiparticle are now the
same particle.
Using this condition, the mass term in the Lagrangian in Eq. (2) can be
written as $\psi^T C^{-1}\psi$, where we have used the fact that $C$ is a
unitary matrix. Writing the mass term in this way makes it clear that if a
field carries a $U(1)$ charge and the theory is invariant under those
$U(1)$ transformations, then the mass term is forbidden. This means that
one cannot impose the Majorana condition on a particle that has a
gauge charge. Since the neutrinos do not have electric
charge, they can be Majorana particles unlike the quarks, electron or the
muon.  It is of course well known that the gauge boson interactions
in a gauge theory Lagrangian conserve a global $U(1)$ symmetry known as
lepton number
with the neutrino and electron carrying the same lepton number. If lepton
number were to be established as an exact symmetry of nature, the Majorana
mass for the neutrino would be forbidden and the neutrino, like the
electron, would be  a Dirac particle.

The properties of a Majorana fermion can be seen in its free field
expansion in terms of creation and annihilation operators:
\begin{eqnarray}
\psi(x) = \int \frac{d^3p}{\sqrt{(2\pi)^3 2E_p}} \Sigma_s \left( a_s({\bf
p})u_s({\bf p}) e^{-ip\cdot x} + a^{\dagger}_s v_s({\bf p}) e^{ip\cdot x}
\right).
\end{eqnarray}
In the gamma matrix convention where $\gamma_i =\left(\begin{array}{cc}
0 & \sigma_i\\ -\sigma_i & 0\end{array}\right)$ and $\gamma_0
=\left(\begin{array}{cc}
0 & {\bf I}\\ {\bf I} & 0\end{array}\right)$, the $u_s$ and $v_s$ are
given by
\begin{eqnarray}
u_s({\bf p}) =\frac{m}{\sqrt{E}}\left(\begin{array}{c}
\alpha_s\\\frac{E-{\bf \sigma\cdot p}}{m}\alpha_s\end{array}\right)
\end{eqnarray}
and
\begin{eqnarray}
v_s({\bf p}) =\frac{m}{\sqrt{E}}\left(\begin{array}{c}
-\frac{E+{\bf \sigma\cdot p}}{m}\alpha'_s\\ \alpha'_s
\end{array}\right).
\end{eqnarray}
$\alpha_s$ and $\alpha'_s$ are two component spinors.

If we choose $\alpha'_s=\sigma_2\alpha_s$, we get
the relation among the spinors $u_s({\bf p})$ and $v_s({\bf p})$
 $C\gamma_0 u^*_s({\bf p}) = v_s({\bf p})$ and the Majorana condition
follows. Note that if $\psi$ were to describe a Dirac spinor, then we
would
have had a different creation operator $b^{\dagger}$ in the second term in
the free field expansion above.

The origin of the two component neutrino is rooted in the isomorphism
between
the Lorentz group and the SL(2, C) group. The latter is defined as the set
of $2\times 2$ complex matrices with unit determinant, whose generators
satisfy the same Lie algebra as that of the LOorentz gtroup. Its basic
representations are $2$ and $2^*$ dimensional. These are the spinor
representations and can be used to describe spin half particles.
 
We can
therefore write the familiar 4-component Dirac spinor used in the text
books to describe an electron can be written as
$\psi=\left(\begin{array}{c} \phi \\ i\sigma_2\chi^*\end{array}\right)$,
where $\chi$ and $\phi$ two two component spinors. 
A Dirac mass is the given by $\chi^T\sigma_2\phi$ whereas a Majorana mass
is given by
$\chi^T\sigma_2 \chi$, where $\sigma_a$ are the Pauli matrices. To make
correspondence with the four component notation, we point out that $\phi$
and $i\sigma_2\chi^*$ are nothing but the $\psi_L$ and $\psi_R$
respectively. It is then clear that $\chi$ and $\phi$ have opposite
electric charges; therefore the Dirac mass $\chi^T\sigma_2\phi$ maintains
electric charge conservation (as well as any other kind of charge like
lepton number etc.).

2-component neutrino is described by the following Lagrangian:
        \begin{eqnarray}
{\cal L}~=~ \nu^\dagger i\sigma^\lambda \partial_\lambda \nu
- {im \over 2} e^{i\delta} \nu^T \sigma_2 \nu
+ {im \over 2} e^{-i\delta} \nu^\dagger \sigma_2 \nu^* \,.
\label{dm2d.lag}
        \end{eqnarray}
This leads to the following equation of motion for the field $\chi$
\begin{eqnarray}
i\sigma^{\lambda}\partial_{\lambda}\chi-im\sigma_2\chi^* = 0
\end{eqnarray}.
As is conventionally done in field theories, we can now give a free field
expansion of the two component Majorana field in terms of the creation and
annihilation operators:
\begin{eqnarray}
\chi(x,t) =\sum_{p,s}[ a_{{\bf p}, s}\alpha_{p,s}e^{-i
p.x}+a^{\dagger}_{{\bf p},s}\beta_{p,s}e^{ip.x}],
\end{eqnarray}
where the sum on $s$ goes over the spin up and down states. 

\noindent{\it Exercise 1}: Using the field equations for a free massive
two
component Majorana spinor, show that its expansion in terms of the
creation
and annihilation operators and two component spinors
$\alpha=\left(\begin{array}{c} 1\\0\end{array}\right)$ and
$\beta=\left(\begin{array}{c} 0\\ 1\end{array}\right)$ is given by the
following expression:
\begin{eqnarray}
\chi(x,t) =\sum_{p}[ a_{{\bf p}, +}e^{-i
p.x}-a^{\dagger}_{{\bf p},-}e^{ip.x}]\alpha \sqrt{E+p}\\ \nonumber
+\sum_{p} [a_{\bf p, -}e^{-p.x} + a^{\dagger}_{\bf p, +}e^{ip.x}]\beta
\sqrt{E-p}.
\end{eqnarray}
Note that in a beta decay process, where a neutron is annihilated and
proton is created, the leptonic weak current that is involved is
$\bar{e}\nu $ (dropping gamma matrices); therefore, along with the
electron, what is created predominantly is a right handed particle
(with a wave function $\alpha$), the
amplitude being of order $\sqrt{E+p}\approx \sqrt{2E}$. This is the
right handed anti 
neutrino. The left handed neutrino is produced
with a much smaller amplitude $\sqrt{E-p}\approx m_nu/E$. Similarly,
in the fusion reaction in the core of the Sun, what is produced is a
left handed state of the neutrino with a very tiny i.e. O($m_\nu/E$)
admixture of the right handed helicity. 

\subsection{Neutrinoless double beta decay and neutrino mass}

As already noted a Majorana neutrino breaks lepton number by two units.
This has the experimentally testable prediction that it leads to the
process of neutrinoless double beta decay, where a nucleus (generally
even-even nuclei) $(Z,N)\rightarrow (Z+2, N-2)+ 2e^-$. We will now show
by using the above property of the Majorana neutrino that if light
neutrino exchange is responsible for this process, then the amplitude is
proportional to the neutrino mass.

Double beta decay involves the change of two neutrons to two protons
and therefore has to be a second order weak interaction process.
 Since each weak
interaction process emits an antineutrino, in second order weak
interaction, the final state will involve two anti-neutrinos. But in
neutrinoless double beta decay, there are no neutrinos in the final state; 
therefore the two neutrinos
must go into the vacuum state. Vacuum state by definition has no spin
whereas the antineutrino emitted in a beta decay has spin. Consider the
antineutrino from one of the decays: it must be predominantly right
handed. But to disappear into vacuum, it must combine with a lefthanded
antineutrino so that the left and right handed spin projections add up to
zero. In the previous paragraph, we showed that the fraction of left
handed spin projection in a neutrino emitted in beta decay is $m_\nu/E$.
Therefore, $\bar{\nu}_e\bar{\nu}_e\rightarrow |0>$ must be proportional to
the neutrino mass. Thus neutrinoless double beta decay is a very sensitive
measure of neutrino mass.

\subsection{Neutrino mass in two component notation}

Let us now discuss the general neutrino mass for Majorana neutrinos. We
saw earlier that for a Majorana neutrinos, there are two different ways to
write a mass term consistent with relativistic invariance.
This richness in the possibility for neutrino masses also has a down side
in the sense that in general, there are more parameters describing the
masses of the neutrinos than those for the quarks and leptons. For
instance 
for the electron and quarks, dynamics (electric charge conservation) 
reduces the number of parameters in their mass matrix. As an example,
using the two component notation for all fermions, for the case of two
two component spinors, a charged fermion mass will be described only by
one parameters whereas for a neutrino, there will be three parameters.
This difference inceases rapidly e.g. for 2N spinors, to describe charged 
fermion masses, we need $N^2$ parameters (ignoring CP violation) whereas
for neutrinos, we need $\frac{2N(2N+1)}{2}$ parameters. What is more
interesting is that for a neutrino like particle, one can have both even
and odd number of two component objects and have a consistent theory.

In this article, we will use two component notation for neutrinos. Thus
when we say that there are N neutrinos, we will mean N two-component
neutrinos. 

In the two component language, all massive neutrinos are Majorana
particles and what is conventionally called a Dirac neutrino is really a
very specific choice of mass parameters for the Majorana neutrino. 
Let us give some examples: If there is only one two component neutrino (we
will drop the prefix two component henceforth), it can have a mass
$m\nu^T\sigma_2\nu$(to be called $\equiv m\nu\nu$ in shorthanded
notation). The neutrino is now a self conjugate object which can be seen
if we write an equivalent 4-component spinor $\psi$:
\begin{eqnarray}
\psi=\left(\begin{array}{c} \nu \\ i\sigma_2\nu^* \end{array}\right)
\end{eqnarray}
Note that this 4-component spinor satisfies the condition 
\begin{eqnarray}
\psi~=~\psi^c\equiv C\bar{\psi}^T
\end{eqnarray}
This condition implies that the neutrino is its own anti-particle, a
 fact more transparent in the 4- rather than the two-component
notation. The above exercise illustrates an important point i.e. given
any two component spinor, one can always write a self conjugate (or
Majorana)
4-component spinor. Whether a particle is really its own antiparticle or
not is therefore determined by its interactions. To see this for the
electrons, one may solve the following excercise i.e. if we wrote two
Majorana spinors using the two two component spinors that describe the
charged
fermion (electron), then until we turn on the electromagnetic
interactions and the mass term, we will not know whether the electron is
its own antiparticle or not. Once we turn on the electromagnetism,
this ambiguity is resolved.

Let us now go one step further and consider two 2-component neutrinos
($\nu_1,~\nu_2$). The general mass matrix for this case is given by:
\begin{eqnarray}
{\cal M}_{2\times 2}~=~\left(\begin{array}{cc} m_1 & m_3 \\ m_3 & m_2
\end{array}\right)
\end{eqnarray}
Note first that this is a symmetric matrix and can be diagonalized by
orthogonal transformations. The eigenstates which will be certain
admixtures of the original neutrinos now describe self conjugate
particles. One can look at some special cases:

\noindent {\it \underline{Case i}:} 

If we have $m_{1,2}=0$ and $m_3\neq 0$, then one can 
assign a charge +1 to $\nu_1$ and -1 to $\nu_2$ under some $U(1)$ symmetry
other than electromagnetism and the theory is invariant under this
extra $U(1)$ symmetry which can be identified as the lepton number and the
particle is then called a Dirac neutrino. The point to be noted is that
the Dirac neutrino is a special case of for two Majorana neutrinos.
In fact if we insisted on calling this case one with two Majorana
neutrinos, then the two will have equal and opposite (in sign) mass as can
be seen
diagonalizing the above mass matrix. Thus a Dirac neutrino can be thought
of as two Majorana neutrinos with equal and opposite (in sign) masses.
Since the argument of a complex mass term in general refers to its C
transformation property (i.e. $\psi^c = e^{i\delta_m}\psi$, where
$\delta_m$ is the phase of the complex mass term), the two two component
fields of a Dirac neutrino have opposite charge conjugation properties.

\noindent {\it \underline{Case ii}:} 

If we have $m_{1,2}\ll m_3$, this case is called
pseudo-Dirac neutrino since this is a slight departure from case (i). In
reality, in this case also the neutrinos are Majorana neutrinos with their
masses $\pm m_0 +\delta$ with $\delta \ll m_0$. The two component
neutrinos will be maximally mixed. Thus this case is of great current
physical interest in view of the atmospheric (and perhaps solar) neutrino
data.

\noindent {\it \underline{ Case iii}:}

 There is third case where one may have $m_1=0$
and $m_3\ll m_2$. In this case the eigenvalues of the neutrino mass matrix
are given respectively by: $m_{\nu}\simeq -\frac{m^2_3}{m_2}$ and $M\simeq
m_2$.
One may wonder under what conditions such a situation may arise in
a realistic gauge model. It turns out that if $\nu_1$ transforms as an
$SU(2)_L$ doublet
and $\nu_2$ is an $SU(2)_L$ singlet, then the value of $m_3$ is limited by
the weak scale whereas $m_2$ has no suchlimit and $m_1=0$ if the theory
has no
$SU(2)_L$ triplet field (as for instance is the case in
the standard model). Choosing $m_2\gg m_3$ then provides a
natural way to understand the
smallness of the neutrino masses. This is known as the seesaw mechanism
\cite{grsyms}. Since this case is very different from the case (i) and
(ii), it is generally said that in grand unified theories, one expects the
neutrinos to be Majorana particles. The reason is that in most grand
unified theories there is a higher scale which under appropriate
situations provides a natural home for the large mass $m_2$. 

While we have so far used only two
neutrinos to exemplify the various cases including the seesaw mechanism,
these discussions generalize when $m_{1,2,3}$ are each $N\times N$
matrices (which we denote by $M_{1,2,3}$). For example, the seesaw formula
for this general situation can
be written as
\begin{eqnarray}
{\cal M}_{\nu} \simeq - M^T_{3D}M^{-1}_{2R} M_{3D}
\end{eqnarray}
where the subscripts $D$ and $R$ are used in anticipation of their origin
in gauge theories where $M_D$ turns out to be the Dirac matrix and $M_R$
is the mass matrix of the right handed neutrinos and all eigenvalues of
$M_R$ are much larger than the elements of $M_D$.
It is also worth pointing out that
Eq. (13) can be written in a more general form where the
Dirac matrices are
not necessarily square matrices but $N\times M$ matrices with $N\neq
M$. We give such examples below.

 Although there is no experimental proof
that the neutrino is a Majorana particle, the general opinion is that the
since the seesaw mechanism provides such a simple way to understand the
glaring
differences between the masses of the neutrinos and the charged fermions
and since it implies that the neutrinos are Majorana fermions, that
they indeed are most likely to be Majorana particles.

Even though for most situations, the neutrino can be treated as a two
component object regardless of whether its mass is of Dirac or Majorana
type, there are certain practical situations where differences between the
Majorana and Dirac neutrino becomes explicit: one case is when the two
neutrinos annihilate. For Dirac neutrinos, the particle and the
antiparticle are distinct and therefore there annihilation is not
restricted by Pauli principle in any manner. However, for the case of
Majorana neutrinos, the identity of neutrinos and antineutrinos plays an
important role and one finds that the annihilation to the Z-bosons occurs
only via the P-waves. Similarly in the decay of the neutrino to any final
state, the decay rate for the Majorana neutrino is a factor of two
higher than for the Dirac neutrino. 

\bigskip

\noindent{\it Exercise 2}: Derive explicit expressions for the four
component 
momentum dependent spinors $u_s(p)$ and $v_s(p)$ when $\psi$ is a Majorana
spinor.

\bigskip

\begin{center} 
{\bf I.3  Experimental indications for neutrino masses}
\end{center}

There have been other lectures at this school on the experimental 
evidences for neutrino masses and their analyses to determine the current
favorite values for the various madd differences as well as mixing angles. 
I will therefore only summarize the main points that are relevant for our
present understanding of neutrino masses and for the sake of
completeness. (For detailed discussion,
see\cite{bahcall} and lectures by Akhmedov, Fogli, Lipari).

At present there is no conclusive evidence for neutrino masses from 
direct search experiments for neutrino masses using tritium beta decay and
neutrinoless double beta decay (see later).

There are however many experiments that have measured the flux of
neutrinos from the Sun and the cosmic rays and which have provided clear
evidences for neutrino oscillations i.e. neutrinos of one flavor
transmuting to neutrinos of another flavor. Since such transmutation can
occur only if the neutrinos have masses and mixings, these experiments
provide evidence for neutrino mass. The expression for vacuum oscillation
probability for neutrinos of a given energy $E$ that have travelled a
distance L is given by:
\begin{eqnarray}
P_{\alpha\beta}~=~\sum_{i,j} |U_{\alpha i}U^*_{\beta i} U^*_{\alpha j}
U_{\beta j}| cos \left(\frac{\Delta_{ij}L}{2E}-\phi_{\alpha \beta, i
j}\right)
\end{eqnarray}
From this it is clear that neutrino oscillation data yields information
about the mass difference squares of the neutrinos ($\Delta_{ij} =
m^2_i-m^2_j$) and mixing angles $U_{\alpha i}$. If the neutrino propagates
in dense matter,
Mikheyev-Smirnov-Wolfenstein effect will change this formula but it also
depends on the same parameters mass difference square and the $U_{\alpha
i}$. The analysis of the data for the atmospheric neutrinos where the
neutrino progates in vacuum and that for solar neutrinos where the effect
of dense matter in the Sun is important therefore leads to the following
picture for masses and mixings:

\noindent{\it Atmospheric neutrinos:}

in the Super-Kamiokande
experiment\cite{atm} which confirms the indications of oscillations in
earlier data from the Kamiokande, 
IMB experiments. More recent data from Soudan II
and MACRO experiments provide further confirmation of this
evidence. The observation here is the following: in the standard model
with massless neutrinos, all the muon and electron
neutrinos produced at the top of the atmosphere would be expected to
reach detectors on the earth and would be isotropic; what has been
observed is that while that is true for the electron neutrinos, the muon
neutrino flux observed on earth exhibit a strong zenith angle dependence.
A simple way to understand this would be to assume that the muon neutrinos
oscillate into another undetected species of neutrino on their way to the
earth, with a characteristic oscillation
length of order of ten thousand kilometers. Since the oscillation length
is roughly given by $E(GeV)/{\Delta m^2(eV^2)}$ kilometers, for a GeV
neutrino, one would expect
the particle physics parameter $\Delta m^2$ corresponding to the mass
difference between the two neutrinos to be around $10^{-3}$ eV$^2$
corresponding to maximal mixing.

From the existing data several important conclusions can be drawn:
(i) the data cannot be fit assuming oscillation between $\nu_{\mu}$ and
$\nu_e$ nor $\nu_\mu-\nu_s$, where $\nu_s$ is a sterile neutrino
which does not any direct weak interaction; (ii) the oscillation scenario
that fit the data best is $\nu_{\mu}-\nu_{\tau}$ for the mass and mixing
parameters
\begin{eqnarray}
\Delta m^2_{\nu_{\mu}-\nu_{\tau}}\simeq (2~-~8)\times 10^{-3}~ eV^2;\\
\nonumber
sin^2 2\theta_{\mu-\tau}\simeq 0.8~-~1
\end{eqnarray}

\noindent{\it Solar neutrinos}

The second evidence for neutrino oscillation comes from the seven
experiments that have observed a deficit in the flux of neutrinos from the
Sun as compared to the predictions of the standard solar model championed
by Bahcall and his collaborators\cite{bahcall} and more recently studied
by many groups. The experiments resposible for this discovery are the
Chlorine, Kamiokande, Gallex, SAGE, Super-Kamiokande, SNO, GNO\cite{solar}
experiments conducted at the Homestake mine, Kamioka in Japan, Gran Sasso
in Italy and Baksan in Russia and Sudbery in Canada. The different
experiments see different
parts of the solar neutrino spectrum. The details of these considerations
are discussed in other lectures.
 The oscillation
interpretation of the solar neutrino deficit has more facets to it than
the atmospheric case: first the final state particle that the $\nu_e$
oscillates into and second what kind of $\Delta m^2$ and mixings fit the
data. At the moment there is a multitude of possibilities. Let us
summarize them now. 

As far as the final state goes, it can either be one of the two remaining
active neutrinos , $\nu_\mu$ and $\nu_{\tau}$ or it can be the sterile
neutrino $\nu_s$. SNO neutral current data announced recently has very
strongly constrained the second possibility (i.e. the sterile neutrino
in the final state). The global analyses of all solar neutrino data seem
to favor the so called large mixing angle MSW solution with parameters:
 $ \Delta m^2 \simeq 1.2\times 10^{-5}-3.1\times 10^{-4} eV^2$;
$sin^22\theta \simeq 0.58~-~0.95$.

\noindent{\it LSND}

Finally, we come to the last indication of neutrino oscillation from the 
Los Alamos Liquid Scintillation Detector (LSND) experiment\cite{lsnd} ,
where neutrino oscillations both from a stopped muon (DAR) as well as the
one accompanying the muon in pion decay
(known as the DIF) have been observed. The evidence from the DAR is
statistically
more significant and is an oscillation from $\bar{\nu}_\mu$ to
$\bar{\nu}_e$. The mass and mixing parameter range that fits data is:
\begin{eqnarray}
LSND: \Delta m^2 \simeq 0.2 - 2 eV^2; sin^22\theta \simeq 0.003-0.03
\end{eqnarray}
There are also points at higher masses specifically at 6 eV$^2$ which are
also allowed by the present LSND data for small mixings. 
KARMEN experiment at the Rutherford laboratory has very strongly
constrained the allowed parameter range of the LSND
data\cite{karmen}. Currently the
Miniboone experiment at Fermilab is under way to probe the LSND parameter
region\cite{louis}.

\noindent{\it Neutrinoless double beta decay and Tritium decay experiment}

Oscillation experiments only depend on the difference of mass squares of
the different neutrinos and the mixing angles. Therefore, in order to have
a complete picture of neutrino masses, we need other experiments. Two such
experiments are the neutrinoless double beta decay searches and the search
for neutrino mass from the analysis of the end point of the electron
energy spectrum in tritium beta decay. 

Neutrinoless double beta decay measures the follwoing combination of
masses and mixing angles:
\begin{eqnarray}
<m>_{\beta\beta}~=~\sum_i U^2_{ei}m_i
\end{eqnarray}
Therefore naively speaking it is sensitive to the overall neutrino mass
scale. But in practice, as we will see below, for the case of both normal
and inverted hierarchies, it is munlikely to settle the question of the
overall mass scale at the presently contempleted level of sensitivity in
double beta decay searches. Only if the neutrino mass patterns are
hierarchical does one expect a visible signal in $\beta\beta_{0\nu}$
decay. We do not get into great details into this issue except to mention
that in drawing any conclusions about neutrino mass from this process,
one has to first have a good calculation of nuclear matrix elements of the
various nuclei involves such as $^{76}$Ge, $^{136}$Xe, $^{100}$Mo etc.;
secondly, another confusing issue has to do with alternative physics
contributions to $\beta\beta_{0\nu}$ which are unrelated to neutrino mass.
Nevertheless, neutrinoless double beta decay is a frundamental experiment
and a nonzero signal will establish a fundamental result that neutrino is
a Majorana particle and that lepton number symmetry is
violated. Regardless of whether it tells us anything about the neutrino
masses, it would provide a fundamental new revelation about physics beyond
the standard model. Presently two experiments Heidelberg-Moscow and IGEX
that use enriched $^{76}$Ge have published limits of $\leq 0.3$
eV\cite{bb}. More
recently, evidence for a double beta signal in the Heidelberg-Moscow data
has been claimed\cite{klap}.

Another important result in further understanding of neutrino mass
physics could come from the tritium end point searches for neutrino
masses. Thisw experiment will measure the parameter $m_\nu =
\sqrt{\sum_i |U_{ei}|^2 m^2_i}$. This involves a different combination of
masses and mixing angles than $<m>_{\beta\beta}$. Presently, the KATRIN
proposal for a high sensitive search for for $m_\nu$ has been made and it
is expected that it can reach a sensitivity of $0.3$ eV.

A third source of information on neutrino mass will come from cosmology,
where more detailed study of structure in the universe is expected to
provide an upper limit on $\sum_i m_i$ of less than an eV.

 Our goal now is to study the theoretical
implications of these
discoveries. We will proceed towards this goal in the following manner: 
we will isolate the mass patterns that fit the above data and then
look for plausible models that can first lead to the general 
feature that neutrinos have tiny masses; then we would try to understand
in simple manner some of the features indicated by data in the hope 
that these general ideas will be part of our final understanding of the
neutrino masses. As mentioned earlier on, to understand the neutrino
masses one has to go
beyond the standard model. First we will sharpen what we mean by this
statement. Then we will present some ideas which may form the basic
framework for constructing the detailed models.

\begin{center}
{\bf I.4  Patterns and textures for neutrinos}
\end{center}

As already mentioned, we will assume two
component neutrinos and therefore their masses will in general be Majorana
type. Let us also give our notation to facilitate further discussion: 
the neutrinos emitted in weak processes such as the beta decay or muon 
decay are weak eigenstates and are not mass eigenstates. The mass
eigenstates 
determine how a neutrino state evolves in time. Similarly, in the
detection process, it is the weak eigenstate that is picked out. This is
of course the key idea behind neutrino oscillation and the
formula presented in the last section. To set the notation, let us
 express the weak eigenstates in terms of the mass
eigenstates. We will denote
the weak eigenstate by the symbol $\alpha, \beta$ or simply $e,\mu, \tau$ 
etc whereas the mass
eigenstate will be denoted by the symbols $i,j,k$ etc. To relate the weak
eigenstates to the mass eigenstates, let us start with the mass terms in
the Lagrangian for the neutrino and the charged leptons:
\begin{eqnarray}
{\cal L}_m~=~\nu^T_L {\cal M}_\nu \nu_L + \bar{E}_L M_\ell E_R + h.c.
\end{eqnarray}
Here the $\nu$ and $E$ which denote the column vectors for neutrinos and
charged leptons are in the weak basis. To go to the mass basis, we
diagonalize these matrices as follows:
\begin{eqnarray}
U^T_L {\cal M}_\nu U_L = d_\nu\\ \nonumber
V_L M_\ell V^{\dagger}_R = d_\ell
\end{eqnarray}
The physical neutrino mixing matrix is then given by:
\begin{eqnarray}
{\bf \Large U}~=~V_L U_L
\end{eqnarray}
 ${\bf \Large U}_{\alpha i}$ and relate the two sets of
eigenstates (weak and mass) as follows:
\begin{eqnarray}
\left(\begin{array}{c} \nu_e\\ \nu_{\mu} \\ \nu_{\tau}\end{array}
\right)=~ U\left(\begin{array}{c} \nu_1 \\ \nu_2 \\ \nu_3
\end{array} \right)
\end{eqnarray}
Using this equation, one can derive the wellknown oscillation formulae for
the survival probability of a particular weak eigenstate $\alpha$
discussed in the previous section.

To see the general structure of the mixing matrix ${\bf \Large U}$, let us
recall that the matrix ${\cal M}_{\nu}$ is complex and symmetric and
therefore has six complex parameters describing it for the case of
three generations. But since the neutrino is described by a complex field,
we can redefine the phases of three fields to remove three
parameters. That leaves nine parameters. In terms of observables, there
are three mass eigenvalues $(m_1,m_2,m_3)$ and three mixing angles and
phases in the mixing matrix ${\bf \Large U}$. The three phases can be
split into one Dirac phase, which is analogous to the phase in the quark
mixing matrix and two Majorana phases. We can then write the matrix 
${\bf \Large U}$ as
\begin{eqnarray}
{\bf \Large U}~=~U^{(0)}\left(\begin{array}{ccc} 1 & &\\ & e^{i\phi_1}& \\
& & e^{i\phi_2}\end{array}\right)
\end{eqnarray}

The matrix ${\bf \Large U^{(0)}}$ has three real angles $\theta_{12},
\theta_{23}, \theta_{13}$ and a phase. The
goal of experiments is to determine all nine of these parameters. The
knowledge of the nine observables allows one to construct the mass matrix
for the neutrinos and from there one can go in search of the new physics
beyond the standard model that leads to such a mass matrix.

The neutrino mass observables given above can be separated into two
classes: (i) oscillation observables and (ii) non-oscillation observables.
The first class of observables are those accessible to neutrino
oscillation experiments and are the two mass differences $\Delta
m^2_{\odot}$ and $\Delta m^2_A$; three mixing angles $\theta_{12}$ (or
$\theta_{\odot}$); $\theta_{23}$ (or $\theta_A$) and $\theta_{13}$ (the
reactor angle, also called $U_{e3}$) and the CP phase $\delta$ in
$U^{(0)}$. The remaining three observables which can only be probed by
nonoscillation experiments are the lightest mass of the three neutrinos
 and the two Majorana phases $\phi_{1,2}$.

\begin{center}
{\bf I.5  Neutrino mixing matrix and mass patterns}
\end{center}

Let us first discuss to what extent the oscillation observables are known.
Our discussion will focus on the three neutrino fits to all neutrino data
which give as central value for $\Delta m^2_A \simeq 0.0025$ eV$^2$; for
solar neutrinos, it gives $\Delta m^2_{\odot} \simeq (2-20)\times 10^{-5}$
eV$^2$. It also provides information on the angles in ${\bf \Large U}$ 
which can be summarized by the
following mixing matrix (neglecting all CP phases):
\begin{eqnarray}
{\bf \Large U}~=~\left(\begin{array}{ccc} c & s & \epsilon \\
-\frac{s+c\epsilon}{\sqrt{2}} & \frac{c-s\epsilon}{\sqrt{2}} &
\frac{1}{\sqrt{2}} \\
-\frac{s-c\epsilon}{\sqrt{2}} & \frac{-c-s\epsilon}{\sqrt{2}} &
\frac{1}{\sqrt{2}} \end{array}\right)
\end{eqnarray}
where $\epsilon \leq 0.16$ from the CHOOZ and PALO-VERDE
reactor experiments\cite{chooz}. $s$ is the solar neutrino mixing angle.
Present experiments allow the range $0.6 \leq sin^2 2\theta_{\odot} \leq
0.96$ with the central being near 0.8. The atmospheric mixing angle
$\theta_A$ is close to maximal i.e. $sin^22\theta_A \simeq 0.8-1$.

As far as the mass pattern goes however, there are three possibilities:
 
\begin{itemize}

\item (i) normal hierarchy: $m_1\ll m_2 \ll m_3$ ; 

\item (ii) inverted hierarchy : $m_1\simeq -m_2 \gg m_3$ and

\item (iii) approximately degenerate pattern \cite{cald1} $m_1\simeq m_2
\simeq m_3$,

\end{itemize}

 where $m_i$ are the eigenvalues of the neutrino mass
matrix. In
the first case, the atmospheric and the solar neutrino data give direct
information on $m_3$ and $m_2$ respectively. On the other hand,
in the last case, the mass differences between the first and
the second eigenvalues will be chosen to fit the solar neutrino data and
the second and the third to fit the atmospheric neutrino data.

Since Majorana masses violate lepton number, a very important
constraint on any discussion of neutrino mass patterns arises from the
negative searches for neutrinoless double beta decay\cite{klapdor}. 
The most stringent present limits are obtained from the Heidelberg-Moscow
enriched Germanium-76 experiment at Gran Sasso and implies an upper limit
on the following combination of masses and mixings:
\begin{eqnarray}
<m_{\nu}>\equiv \Sigma_i U^2_{ei} m_{\nu_i} \leq 0.35 ~eV~~~~~~95\%~c.l.
\end{eqnarray}
This upper limit depends on the nuclear matrix element calculated by the
Heidelberg group\cite{klapdor}. There could be an uncertainty of a factor
of two in this estimate. This would then relax the above upper bound
to at most 0.7 eV in the worst case scenario . This is still a very useful
limit and becomes especially
relevant when one considers whether the neutrinos constitute a significant
fraction of the hot dark matter of the universe. A useful working formula
is $\Sigma_i m_{\nu_i}\simeq 24\Omega_{\nu} $ eV where $\Omega_{\nu}$ is
the neutrino fraction that contributes to the dark matter of the universe.
For instance, if the neutrino contribution to dark matter fraction is
20\%, then the sum total of
neutrino masses must be 4.8 eV. Such large values are apparently in
disagreement with present upper limits from structure surveys such as 2dF
survey and others.

Neutrinoless double beta decay limits also imply very stringent 
constraints on the mixing pattern in the degenerate case.

The inverted hierarchy case  (ii) is quite an interesting one and will be
discussed from a theoretical perspective in more detail later on; but at
the moment we simply note that in this case the value of $m_1$ is nothing
but the $\sqrt{\Delta m^2_A}$. The solar mass difference is an additional
parameter in the mass matrix.

At this point, it is appropriate to stress the theoretical challenges
raised by the existing neutrino oscillation data.

\begin{itemize}

\item {\it Ultralight neutrinos:} Why are the neutrino masses so much
lighter than the quark and charged lepton masses ?

\item {\it Near bimaximal mixing:} How to understand simultaneously two
large mixing angles one for the $\mu-\tau$ and another for $e-\mu$ ?

\item {\it Smallness of $\Delta m^2_{\odot}/\Delta m^2_A$ :}
Experimentally, $\Delta m^2_{\odot}\simeq 10^{-2} \Delta m^2_A$. How does
one understand this in a natural manner ?

\item {\it Smallness of $U_{e3}$:} The reactor results also seem to
indicate that the angle $\theta_{13} \equiv U_{e3}$ is a very small
number. One must also understand this in a framework that simultanoeusly
explains all other puzzles.

\end{itemize}

 Possible other puzzles include a proper understanding of neutrino mass
degeneracy if there is a large positive signal for the neutrinoless double
beta decay and of course, when we have evidence for CP violating phases in
the mass matrix, we must understand their magnitude.

\bigskip

\begin{center}
{\bf I.6~~Neutrino mass textures}
\end{center}

From the mixing matrix in Eq. (24), we can write down the allowed neutrino
mass matrix for any arbitrary mass pattern assuming the neutrino 
is a Majorana fermion. Denoting the matrix elements
of ${\cal M}_\nu$ as $\mu_{\alpha\beta}$ for $\alpha, \beta = 1,2,3$, we
have(Recall that $\mu_{\alpha \beta} = \mu_{\beta \alpha}$):
 \begin{eqnarray}
\mu_{11} = [{c^2}{m_1}+{s^2}{m_2}+{\epsilon^2}{m_3}] \\
\nonumber
\mu_{12} =\frac{1}{\sqrt{2}}[ -{c(s+ c\epsilon)}{m_1} +{s(c-
s\epsilon)}{m_2} +{\epsilon}{m_3}]
\\ \nonumber
\mu_{13} =\frac{1}{\sqrt{2}}[ -{c(s- c\epsilon)}{m_1} - {s(c+
s\epsilon)}{m_2} +{\epsilon}{m_3}] \\ \nonumber
\mu_{22} =\frac{1}{2}[ {(s+c\epsilon)^2}{m_1} +{(c-
s\epsilon)^2}{m_2} + {m_3}] \\ \nonumber
\mu_{23} =\frac{1}{2}[-{(s^2- c^2\epsilon^2)}{m_1}-{(c^2-
s^2\epsilon^2)}{m_2} +{m_3}] \\ \nonumber
\mu_{33} =\frac{1}{2} [{(s- c\epsilon)^2}{m_1}+{(c+
s\epsilon)^2}{m_2} +{m_3}].
\end{eqnarray}

One can use Eq. (26) to get information on the nature of the neutrino mass
matrix and use it to get clues to the nature of physics beyond the
standard model.
One technique is to rewrite this mass matrix in a
way that reflects some new underlying symmetry of physics beyond the
standard model. Below we give some examples of mass matrices that are
closely related to this mass matrix but reflect some symmetries of the
lepton world. A second utility of Eq. (26) is to search for mass matrices
that lead to testable predictions and thereby test models.

As an example consider the case when $\mu_{11}=\mu_{22}=\mu_{33}=0$. This
is the prediction from the Zee model\cite{zee}. Using
Eq. (26), one can
easily see that in the leading order the vanishing of all diagonal entries
implies that $\Delta m^2_{odot}=0$ i.e. one must keep higher order terms
in $\epsilon$ to get a nonzero $\Delta m^2_{\odot}$.

An elementary way to proceed in this directions is to start with ways to
understand the maximal mixing using only two flavors as in the case of the 
atmospheric neutrinos (i.e. the 2-3
sector of the neutrino flavor). It is well known that if we have a matrix
of the form
\begin{eqnarray}
M~=~\pmatrix { A & B \cr B & A}
\end{eqnarray}
then, its eigenstates are maximal admixtures of the original ``flavor''
states. The eigen values are $(A+B), (A-B)$. It is now clear that if we
want one of the eigenvalues to be much less than the other, we must have
$A\approx B$. We will now have to generalize this discussion to the case
of three generations. For this we note that, the ``mixing matrix'' for
this case can be written as:
\begin{eqnarray}
U~=~\pmatrix{1 & 1\cr 1 & -1}
\end{eqnarray}
We will call this matrix ``maximal mixing matrix''.

\bigskip

\noindent{\it Theorem:}
The maximal mixing matrix for N-generations, if N is a prime number is
given by:
\begin{eqnarray}
U_N~=~\frac{1}{\sqrt{N}}\pmatrix{1 & 1 & 1 & \cdot & \cdot \cr
1 & z & z^2 & z^3 & \cdot \cr 1 & z^2 & z^4 & \cdot & \cdot \cr
1 & z^3 & z^6 & \cdot &\cdot \cr 1 & z^4 & z^8 & \cdot & \cdot }
\end{eqnarray}
where $z$ is the N-th root of unity and the rows and extend in an obvious
manner to make the matrix $N\times N$.
Note that the general form of an arbitrary element is $z^{pq}$. When $pq
\geq N$, the power is simply given by $pq-N$.
When the number of generations is not a prime number, then $N=N_1\cdot N_2 
\cdot N_3 \cdot\cdot$. In this case,
\begin{eqnarray}
U_N~=~ U_{N_1}\times U_{N_2} \times U_{N_3} \times \cdot 
\end{eqnarray}
where $\times$ stands for a direct product. For example for the case of
$N=4$, we have for the maximal mixing matrix $U_4~=~ U_2 \times U_2$ which
is given by
\begin{eqnarray}
U_4~=~\frac{1}{2}\pmatrix{1 & 1&1&1\cr 1&-1&1&-1 \cr 1&1&-1&-1 \cr 1&-1
&-1 &1}
\end{eqnarray}
The four by four case turns out to be interesting since by decoupling
a linear combination of the states from the $4\times 4$ system, we can get
the exact bimaximal form\cite{nussinov}. The corresponding $3\times 3$
mass matrix can also be gotten by starting from this discussion. To see
how one gets that, note that the matrix in Eq.(24) for the $2\times 2$
case has a $S_{12}$ symmetry. The corresponding symmetry in the $4\times
4$ case is then $S_{12}\times S_{34}$. Using this property and decoupling
one linear combination of the fields, we can get the mass matrix for the
$3\times 3$ case that leads to bimaximal mixing. It is left as an exercise
to the reader to show that the corresponding mass matrix is given by:
\cite{nussinov}:
 \begin{eqnarray}
M_{\nu}=\left(\begin{array}{ccc} A+D & F & F \\ F & A & D \\ F & D & A
\end{array}\right)
\end{eqnarray}

Since the present data implies that there are deviations from the exact
bimaximal form, this mass matrix must only be treated as a leading order 
contribution.

The three different mass patterns can emerge from this mass matrix in
various limits: e.g. (i) for $F \ll A \simeq -D$, one gets the normal
hierarchy; (ii) for $F \gg A, D$, one has the inverted pattern for masses
and (iii) the parameter region $F, D \ll A$ leads to the degenerate case.
Clearly this mass matrix is the leading order matrix and there will have
to be small corrections to this to fit solar neutrino data as well as any
possible evidence (or lack of it) for neutrinoless double beta decay. An
interesting symmetry of this mass matrix is the $\nu_\mu \leftrightarrow
\nu_\tau$ interchange symmetry, which is obvious from the matrix; but in
the limit where $A=D=0$, there appears a much more interesting symmetry
i.e. the continuous symmetry $L_e-L_\mu-L_\tau$\cite{emutau}. If the
inverted mass matrix is confirmed by future experiments, this symmetry
will provide an important clue to new neutrino related physics beyond the
standard model (for an example see \cite{kuchi}). 

There are other ways to proceed towards the same goal. One way inspired by
the studies of quark mass matrices is to consider mass matrices with zeros
in it and hope that there are sensible symmetries that will guarantee the
zeros. One may then be able to obtain relations between different
observables such as masses and mixing angles that can be tested in
experiments.

To proceed towards this goal, let us recall that in the absence of CP
violation, the neutrino mass matrix has six independent entries. In terms
of observables there are five oscillation observables two mass differences 
$(\Delta m^2_{\odot}, \Delta m^2_A)$, and three mixing
angles solar, atmospheric and the reactor angle $U_{e3}$.
 So if we have more than one vanishing element, we will have
nontrivial relations between oscillation observables which can be used as
tests of the various ansatzes.

First point to note is that if there are three zeros or more, the mass
matrix cannot describe data. The proof of this is left as an exercise to
the reader. 

\bigskip
\noindent{\it Exercise} Show that the following ``three zero'' mass matrix
$M_{3-zero}~=~\pmatrix{0 & a & b\cr a & 0 & c\cr b & c & 0}$ predicts
$sin^22\theta_{\odot} =1-\frac{1}{16}\frac{\Delta m^2_{\odot}}{\Delta
m^2_A}$ and is therefore ruled out by present neutrino data.

Let us therefore consider some typical two zero mass
matrices\cite{frampton}:
\begin{eqnarray}
M_{\nu}=\sqrt{\Delta m^2_A}\left(\begin{array}{ccc} 0 & 0 &d\epsilon \\ 0
& 1+ a\epsilon & 1 \\ d\epsilon & 1 & 1+b\epsilon
\end{array}\right)
\label{ms}
\end{eqnarray}
or alternatively
\begin{eqnarray}
M_{\nu}=\sqrt{\Delta m^2_A}\left(\begin{array}{ccc} 0 & d\epsilon & 0 \\
d\epsilon
& 1+ a\epsilon & 1 \\ 0 & 1 & 1+b\epsilon
\end{array}\right)
\end{eqnarray}
They gives a four parameter fit to all data and yield a relation between
$\Delta m^2_{\odot}/\Delta m^2_A$, $U_{e3}$, $\theta_{\odot}$ and
$\theta_A$\cite{frampton}.
\begin{eqnarray}
U^2_{e3}cos2\theta_{\odot}~=~\frac{sin^22\theta_{\odot}}{4}\frac{\Delta
m^2_{\odot}}{\Delta m^2_A}
\end{eqnarray}
This relation predicts that $U_{e3}\geq 0.12$ and can be tested in planned
long baseline experiments to be conducted in near future.

There are also some other two zero textures that predict a degenerate
spectrum for neutrinos.

So far we have focussed on studying the form of the neutrino mass
matrix. The implicit assumption in these discussions has been that the
charged lepton mass matrix is diagonal. It may however be that all
neutrino mixings are a reflection of structure in the charged lepton
mass matrix rather than in the neutrinos. An example of this kind is
 the democratic mass matrix\cite{xing}
which also leads to a near bimaximal mixing.
As was noted in Ref.\cite{nuss2}, in order to get the democratic mixing
matrix,
one must choose the charged lepton mass matrix in the following form while
keeping
the neutrino mass matrix diagonal.:
\begin{eqnarray}
M_{\ell}=\left(\begin{array}{ccc} a & 1 & 1 \\ 1 & a & 1 \\ 1 & 1 & a
\end{array}\right)
\end{eqnarray}
This form can be derived from a permutation symmetry $S_3$ operating on
the lepton doublets, which may provide another clue to possible gauge
model building.

Other models that use charged lepton sector to understand neutrino has
been the focus of
several papers\cite{barr},specially in the context of grand unified
theories\cite{barr}. We do not discuss this class of models here.

\begin{center}
{\bf I.7:  Inverted hierarchy
and $L_e-L_\mu-L_\tau$ symmetry }
\end{center}

In this section, we consider another interesting clue to model building
present in neutrino data if the mass arrangement is inverted. As already
noted, if in eq.\ref{ms},
we set $A=D=0$, this leads to a neutrino mass matrix with two degenerate
neutrinos with mass $\pm\sqrt{2}F$ and one massless neutrino. The
atmospheric mass difference is given by $\Delta m^2_A = 2F^2$ and mixing
angle $\theta_A=\pi/4$. As far as the solar $\nu_e$ oscillation is
concerned, the $sin^22\theta_{\odot}=1$ but $\Delta m^2_{\odot}=0$. While
this is unphysical, this raises the hope that as corrections to this mass
matrix are taken into account, it may be possible understand the smallness
of $\Delta m^2_{\odot}/\Delta m^2_A$ naturally.

In fact this hope is fortified by the observation that the $A=D=0$ limit
of the mass matrix in ref.\ref{ms} has the leptonic symmetry
$L_e-L_\mu-L_\tau$; therefore one might hope that as this symmetry is
broken by small terms, one will end up with a situation that fits data
well.

This question was studied in two papers\cite{babu,goh}. To proceed with
the discussion, let us consider the following mass matrix for neutrinos
where small $L_e-L_\mu-L_\tau$ violating terms have been added.
\begin{eqnarray}
{\cal M}_\nu=m~\left(\begin{array}{ccc} z &
c & s\\ c & y & d\\ s & d & x\end{array}\right).
\end{eqnarray}
The charged lepton mass matrix is chosen to have a diagonal form in this
basis and $L_e-L_\mu-L_\tau$ symmetric.

 In the perturbative
approximation, we find the following sumrules
involving the neutrino observables and the elements of the neutrino mass
matrix. The two obvious relations are
 \begin{eqnarray}
\sin^22\theta_{A}~=~\sin^22\theta + O(\delta^2)\cr
D_3 \equiv \triangle m_A^2~=~-m^2 + 2 \triangle m^2_{\odot}~+~
O(\delta^2)\cr
\end{eqnarray}
The nontrivial relations that also hold for this model are:
\begin{eqnarray}
\sin^22\theta_{\odot}~=~1-(\frac{\triangle m_\odot^2}{4\triangle
m_A^2}-z)^2~+~O(\delta^3) \cr
 \frac{\triangle m_\odot^2}{\triangle
m_A^2}~=~2(z+\vec{v}\cdot\vec{x})~+~O(\delta^2)\cr
U_{e3}~=~\vec{A}\cdot(\vec{v}\times\vec{x})~+~O(\delta^3)\cr
\end{eqnarray}
where $\vec{v}=(\cos^2\theta,\sin^2\theta,\sqrt{2}\sin\theta\cos\theta)$,
$\vec{x}=(x,y,\sqrt{2}d)$ and
$\vec{A}~=~\frac{1}{\sqrt{2}}(1,1,0)$. $\delta$
in the preceding equations represents the
small parameters in the mass matrix. These equations represent one of the
main results of this paper. Below we study their implications. Finally,
there is the relation $\langle m \rangle_{\beta\beta}~=~m z$. This is an
an exact relation true to all orders in the small parameters. 

One of the major consequences of these relations is that (i) there is a
close connection between the measured value of the solar mixing angle and
the neutrino mass measured in neutrinoless double beta decay; (ii) the
present values for the solar mixing angle can be used to predict the
$m_{\beta\beta}$ for a value of the $\Delta m^2_{\odot}$. For instance, 
for $sin^22\theta_{\odot}=0.9$, we would predict 
$(\frac{\triangle m_\odot^2}{4\triangle m_A^2}-z) =0.3$. For small $\Delta
m^2_{\odot}$, this implies $m_{\beta\beta}\simeq 0.01$ eV. 
The second relation involving the $\Delta m^2_{\odot}/\Delta m^2_A$ in 
terms of $x, y , z, d$ tells us that for this to be the case, we must
have strong cancellation between the various small parameters. Given
this, the above $m_{\beta\beta}$ value
is expected to be witihin the reach of new double beta decay experiments
contemplated\cite{dbeta}. Note however that the $sin^22\theta_{\odot}$
cannot be larger than $0.9$ in the case of approximate
$L_e-L_\mu-L_\tau$ symmetry.

If the value of  $sin^22\theta_{\odot}$ is ultimately determined to be
less than $0.9$, the question one may ask is whether the idea of
$L_e-L_\mu-L_\tau$ symmetry is dead. The answer is in the negative since
so far we have explored the breaking of $L_e-L_\mu-L_\tau$ symmetry only
in the neutrino mass matrix. It was shown in \cite{babu} that if the
symmetry is broken in the charged lepton mass, one can lower the
$sin^22\theta_{\odot}$ as long as the value of $U_{e3}$ is
sizable. However given the present upper limit on $U_{e3}$, the smallest
value is somewhere around $sin^22\theta_{\odot}\simeq 0.8$.

\begin{center}
{\bf I.8  CP violation}
\end{center}

A not very well explored aspect of neutrino physics at the moment is CP
violation in lepton physics. Unlike the quark sector, CP violation for
Majorana neutrinos allows for more phases for neutrinos. Since the
Majorana neutrino mass matrix is symmetric, for $N$
generations of neutrinos, there are in general
$\frac{N(N+1)}{2}$ phases in it. When the mass matrix is diagonalized,
these phases will appear in the unitary matrix $U_L$ that does the
diagonalization (i.e. $U^T-{\cal M}_\nu U_L = d_\nu$). If we are working
in a basis where the charged lepton
mass matrix is diagonal, then $U_L$ leptonic weak mixing matrix. As we
saw this has $N(N+1)/2$ phases. Out of them,
redefinition of the charged lepton fields in the weak current allows the
removal of N phases; so there are $N(N-1)/2$ phases in the neutrino
masses. In the quark sector, both up and down fields could be redefined
allowing for the number of physical phases that appear in the end to be
smaller. However for Majorana neutrinos, redefinition of the fields does
not remove the phases entirely from the theory but rather shifts them to
other places where they can manifest themselves physically\cite{boris}.

Thus for two generations there is one and for three generations
there are 3 phases. A convenient parameterization of the mixing matrix
with these phases in the mixing matrix for the case of three generations 
is ${\bf \Large U}{\bf \Large K}$:
\begin{eqnarray}
{\bf \Large U}~=~\left(\begin{array}{ccc} c & s & \epsilon e^{-i\delta}\\
-\frac{s+c\epsilon e^{i\delta}}{\sqrt{2}} & \frac{c-s\epsilon
e^{i\delta}}{\sqrt{2}} &
\frac{1}{\sqrt{2}} \\
-\frac{s-c\epsilon e^{i\delta}}{\sqrt{2}} & \frac{-c-s\epsilon
e^{i\delta}}{\sqrt{2}} &
\frac{1}{\sqrt{2}} \end{array}\right)
\end{eqnarray}
and 
\begin{eqnarray}
{\bf \Large K} ~=~ Diag (1, e^{i\phi_1}, e^{i\phi_2})
\end{eqnarray}
The phase $\delta$ is called the Dirac phase and the $\phi_i$ are called
Majorana phases. Note that the Dirac phase is always multiplied by the
small mixing angle $\epsilon$. Its measurability is therefore very closely
tied to the absolute magnitude of $\epsilon$. Coming to the Majorana
phases, one of the two phases
$\phi_i$ can in principle be probed in neutrinoless  double beta decay.
To see this, let us note that
\begin{eqnarray}
<m>_{\beta\beta}~=~|m_1 c^2 + m_2 s^2e^{i\phi_1}+m_3 \epsilon^2
e^{-i(\delta-\phi_2)}|
\end{eqnarray}
Since $\epsilon\ll 1$, the last term in the above equation can be dropped.
It is then easy to see that one has some chance of seeing the CP phase
$\phi_1$, once one has a precise knowledge of the $\Delta m^2_A$ and the
solar mixing angle, provided the nuclear matrix elements are known
better\cite{smirnov}. More optimistically, when the phase is zero, the two
terms in the expression for $\beta\beta_{0\nu}$ add up and the chances for
seeing it is enhanced.

 On the whole
though, there is some chance of measuring the CP phase both for the
inverted and the degenerate mass case provided the nuclear matrix
elements have much smaller uncertainties than presently known, whereas for
the case of normal
hierarchy, it depends on how small the smallest neutrino mass is. If it is
very close to $m_2$, the so called quasi-degenerate case, then one has a
good chance to measure the phase if the $<m>_{\beta\beta}$ is measured to
a precision of 0.001 eV.

A very interesting question relates to the observability of truly CP
violating leptonic processes. This question has been addressed in
\cite{andre}. For instance although one can probe the CP phase, it is not
really a CP violating process; in other words, even in the presence of CP
violation, the rate for neutrinoless double beta decay of a nucleus is
same as that for the corresponding anti-nucleus. On the other hand, there
are genuine CP violating processes where the CP phase can be probed. One
is the celebrated example of early universe leptogenesis where one looks
at the decay of the Majorana right handed neutrino into $\ell +H$ and
$\bar{\ell} +\bar{H}$ and it is their difference that manifests as the
lepton asymmetry.  Similarly, one can look at rare decays of $K^\pm$ to
$\pi^\mp + \mu^\pm + \mu^\pm$ and similar decay modes for the $B$ meson,
where the presence of a physical intermediate state leads bto the
observability of a truly CP violating difference between decay rates.

\newpage

\section{}

\begin{center}
{\bf II.1  Why neutrino mass requires physics beyond
the standard model ?}
\end{center}

We will now show that in the standard model, the neutrino mass vanishes 
to all orders in perturbation theory as well as nonperturbatively. The
standard model is based on the gauge group
$SU(3)_c\times SU(2)_L\times U(1)_Y$ group under which the quarks and
leptons transform as described in the Table I.

\begin{center}
{\bf Table I}
\end{center}

\begin{center}
\begin{tabular}{|c||c|}
\hline\hline
 Field &  gauge  transformation \\ \hline\hline
 Quarks $Q_L$ & $(3,2, {1\over 3})$\\
 Righthanded up quarks $u_R$ &  $(3, 1, {4\over 3})$ \\
Righthanded down quarks  $ d_R$ &  $(3, 1,-\frac{2}{3})$\\
Lefthanded  Leptons $L$ & $(1, 2 -1)$ \\
 Righthanded leptons  $e_R$ & $(1,1,-2)$ \\
Higgs Boson $\bf H$ & $(1, 2, +1)$ \\
Color Gauge Fields  $G_a$ & $(8, 1, 0)$ \\
Weak Gauge Fields  $W^{\pm}$, $Z$, $\gamma$ & $(1,3+1,0)$ \\
\hline\hline
\end{tabular}
\end{center}

\noindent {\bf Table caption:} The assignment of particles to the standard
model gauge group $SU(3)_c\times SU(2)_L\times U(1)_Y$.

The electroweak symmetry $SU(2)_L\times U(1)_Y$ is broken by the vacuum
expectation of the Higgs doublet $<H^0>=v_{wk}\simeq 246$ GeV, which gives
mass to the gauge bosons and the fermions, all fermions except the
neutrino. Thus the neutrino is massless in the standard model, at the tree
level. There are several questions that arise at this stage. What happens
when one goes beyond the above simple tree level approximation ? Secondly,
do nonperturbative effects change this tree level result ? Finally, how to
judge how this result will be modified when the quantum gravity effects
are included ?

The first and second questions are easily answered by using the B-L
symmetry of the standard model. The point is that since the standard model
has no $SU(2)_L$ singlet neutrino-like field, the only possible mass terms
that are allowed by Lorentz invariance are of the form
$\nu^T_{iL}C^{-1}\nu_{jL}$, where $i,j$ stand for the generation index and
$C$ is the Lorentz charge conjugation matrix. Since the $\nu_{iL}$ is part
of the $SU(2)_L$ doublet field and has lepton number +1, the above
neutrino mass term transforms as an $SU(2)_L$ triplet and furthermore, it
violates total lepton number (defined as $L\equiv L_e+L_{\mu}+L_{\tau}$)
by two units. However, a quick look at the standard model Lagrangian
convinces one that the model has exact lepton number symmetry after
symmetry breaking; therefore such terms can never arise in perturbation
theory.
Thus to all orders in perturbation theory, the neutrinos are massless.
As far as the nonperturbative effects go, the only known source is the
weak instanton effects. Such effects could effect the result if they
broke the lepton number symmetry. One way to see if such breaking
weak instanton effects. Such effects could effect the result if they
broke the lepton number symmetry. One way to see if such breaking
occurs is to look for anomalies in lepton number current conservation from
triangle diagrams. Indeed $\partial_{\mu}j^{\mu}_{\ell}= c W \tilde{W} +
c' B\tilde{B}$ due to the contribution of the leptons to the triangle
involving the lepton number current and $W$'s or $B$'s. Luckily, it turns
out that the anomaly contribution to the baryon number current
nonconservation has also an identical form, so that the $B-L$ current
$j^{\mu}_{B-L}$ is conserved to all orders in the gauge couplings. As a
consequence, nonperturbative effects from the gauge sector cannot induce
$B-L$ violation. Since the neutrino mass operator described above violates
also $B-L$, this proves that neutrino masses remain zero even in the
presence of nonperturbative effects.

Let us now turn to the effect of gravity. Clearly as long as we treat
gravity in perturbation theory, the above symmetry arguments hold since
all gravity coupling respect $B-L$ symmetry. However, once nonperturbative
gravitational effects e.g black holes and worm holes are
included, there is no guarantee that global symmetries will
be respected in the low energy theory. The intuitive way to appreciate the
argument is to note that throwing baryons into a black hole does not lead
to any detectable consequence except thru a net change in the baryon
number of the universe. Since one can throw in an arbitrary numnber of
baryons into the black hole, an arbitrary information loss about the net
number of missing baryons would prevent us from defining a baryon
number of the visible
universe- thus baryon number in the presence of a black hole can not be an
exact symmetry. Similar arguments can be made for any global charge such
as lepton number in the standard model. A field theoretic parameterization 
of this statement is that the effective low energy Lagrangian for the
standard model in the presence of black holes and worm holes etc must
contain baryon and lepton number violating terms. In the context of the
standard model, the only such terms that one can construct are
nonrenormalizable terms of the form $~LH LH/M_{P\ell}$. After gauge
symmetry breaking, they lead to neutrino masses; however these masses are
at most of order $~v^2_{wk}/M_{P\ell}\simeq 10^{-5}$ eV\cite{akm}.
But
as we discussed in the previous section, in order to solve the atmospheric
neutrino problem, one needs masses at least three orders of magnitude
higher.

Thus one must seek physics beyond the standard model to explain observed
evidences for neutrino masses. While there are many possibilities that
lead to small neutrino masses of both Majorana as well as Dirac kind, here
we focus on the possibility that there is a heavy right handed
neutrino (or neutrinos) that  lead to a small 
neutrino mass. The resulting
mechanism is known as the seesaw mechanism and leads to neutrino being a
Majorana particle.

The nature and origin of the seesaw mechanism can also be tested in other
experiments and we will discuss them below. This will be dependent on the
kind of operators that play a role in generating neutrino masses. If the
leading order operator is of dimension 5, then the scale necessarily is
very high (of order $10^{12}$ GeV or greater). On the other hand, in
theories with extra space dimensions, this operator may be forbidden and
one may be forced to go to higher dimensional operators, in which case the
scale could be lower. In the lecture IV, an example of five and six
dimensional theory is given, where this indeed happens.

The seesaw mechanism raises a very important question: since we require
the mass of the right handed neutrino to be much less than the Planck
scale, a key question is ``what symmetry keeps the right handed neutrino
mass lighter?'' We will give two examples of symmetries that can do this.

 \bigskip 

\begin{center}{\bf II.2~~ Seesaw and the right handed neutrino}
\end{center}

 The simplest possibility extension of the standard model that leads to
nonzero mass for the neutrino is one where only a right handed neutrino is
added to the standard model. In this case $\nu_L$ and $\nu_R$ can
form a mass term; but apriori, this mass term is like the mass terms for
charged leptons or quark masses and will therefore involve the weak scale.
If we call the corresponding Yukawa coupling to be $Y_\nu$, then the
neutrino mass is $m_D=Y_\nu v/\sqrt{2}$. For a neutrino mass in the eV
range
requires that $Y_\nu \simeq 10^{-11}$ or less. Introduction of such small
coupling constants into a theory is generally considered unnatural and a
sound theory must find
a symmetry reason for such smallness. As already already
alluded to before, seesaw mechanism\cite{grsyms}, where we introduce a
singlet Majorana mass term for the right handed neutrino is one way to
achieve this goal. What we have in this case is a
$(\nu_L,\nu_R)$ mass matrix which has the form: 
\begin{eqnarray}
M=\left(\begin{array}{cc} 0 & m_D \\
m_D & M_R\end{array}\right)
\end{eqnarray}
Since $M_R$ is not constrained by the standard model symmetries, it is
natural to choose it to be at a scale much higher than the weak scale.
 Now diagonalizing this mass matrix for a single neutrino species, we get
a heavy eigenstate $N_R$ with mass $M_R$ and a light
eigenstate $\nu$ with mass $m_\nu\simeq \simeq \frac{-m^2\nu}{M_{R}}$.
This provides a natural way to understand a small neutrino mass without
any unnatural adjustment of parameters of a theory. In a subsequent
section, we will discuss a theory which connects the scale $M_R$ to a new
symmetry of nature beyond the standard model.

\begin{center}
{\bf II.2A Why is $M_{\nu_R}\ll M_{P\ell}$  ?}
\end{center}

The question ``why $M_{\nu_R}\ll M_{P\ell}$ ?'' is in many ways similar to 
the question in the standard model i.s. ``why is $M_{Higgs} \ll M_{P\ell}$
?'' It is well known that searches for answer to this question has led us
to consider many interesting possibilities for physics beyond the standard
model and supersymmetry appears to be the most promising answer to this
question. It is hoped that answering this question for $\nu_R$ can also
lead us to new insight into new symmetries beyond the standard model.
There are two interesting answers to our question that I will elaborate
later on.

\bigskip

\noindent {\bf \Large $B-L$}:

\bigskip

If one adds three right handed neutrinos to implement the seesaw
mechanism, the model admits an anomaly free new symmetry i.e. $B-L$. 
One can therefore extend the standard model symmetry to either
$SU(2)_L\times U(1)_{I_{3R}}\times U(1)_{B-L}$ or its left-right
symmetric extension $SU(2)_L\times SU(2)_R\times U(1)_{B-L}$. In either
case the right handed neutrino carries the B-L quamtum number and its
Majorana mass breaks this symmetry. Therefore, the mass of the $\nu_R$ can
at most be the scale of $B-L$ symmetry breaking, hence answering the
question ``why $M_{\nu_R}\ll M_{P\ell}$ ?''.

\bigskip

\noindent {\bf\Large $SU(2)_H$:} 

\bigskip

While local $B-L$ is perhaps the most straight forward and natural
symmetry that keeps $\nu_R$ lighter than the Planck scale, another
possibility has recently been suggested in ref. \cite{kuchi}. The
mainobservation here is that is the standard model is extended by
including a local $SU(2)_H$ symmetry acting on the first two lepton
generations in cluding the right handed charged leptons, then global
Witten anomaly freedom dictates that there must be at least two right
handed neutrinos which transform as a doublet under the $SU(2)_H$ local
symmetry. In this class of models, in the limit of exact $SU(2)_H$
symmetry, the $\nu_R$'s are massless and as soon as the $SU(2)_H$ symmetry
is broken, they pick up mass. Therefore ``lightness'' of the $\nu_R$'s
compared to the Planck scale in these models is related to an $SU(2)_H$
symmetry. These comments are elaborated with explicit examples later on in
this review.

\begin{center}
{\bf II.2B  Small neutrino mass using a double seesaw mechanism with 
$\nu_R$}
\end{center}

As we saw from the previous discussion, the conventional seesaw mechanism
requires rather high mass for the right handed neutrino and therefore
a correspondingly high scale for B-L symmetry breaking.
 There is however no way at
present to know what the scale of B-L symmetry breaking is. There are
for example models bases on string compactification\cite{langacker} where
the $B-L$ is quite possibly is in the TeV range. In this case small
neutrino mass can be implemented by a double seesaw mechanism suggested in
Ref.\cite{valle}. The idea is to take a right handed neutrino $N$ and a
singlet neutrino $S$ which has extra quantum numbers which prevent it from
coupling to the left handed neutrino. One can then write a three by three
neutrino mass matrix in the basis $(\nu, N, S)$ of the form:
\begin{eqnarray}
{ M}~=~\pmatrix{0 & m_D & 0 \cr m_D & 0 & M \cr 0 & M
& \mu}
\end{eqnarray}
For the case $\mu \ll M \approx M_{B-L}$, (where $M_{B-L}$ is the $B-L$
breaking scale) this matrix has one light and two heavy
neutrinos per generation and the latter two form a pseudo-Dirac pair
with mass of order $M_{B-L}$. The important thing
for us is that the light mass eigenvalue is given by $m^2_D\mu/M^2$; for
$m_D\approx \mu \simeq $ GeV, a 10 TeV $B-L$ scale is enough to give
neutrino masses in the eV range. For the case of three generations, the
formula for the light neutrino mass matrix is given by:
\begin{eqnarray}
{\cal M}_\nu~=~M_DM^{-1}\mu M^{-1}M^T_D
\end{eqnarray}

\begin{center}
{\bf II.3  High mass Higgs triplet induced neutrino masses}
\end{center}

As already discussed, one way to generate nonzero neutrino masses 
without using the righthanded neutrino is to include in the standard model 
an $SU(2)_L$ triplet Higgs field with $Y=2$ so that the electric charge
profile of the members of the multiplet is given as follows:
$(\Delta^{++}, \Delta^{+}, \Delta^0)$.  This allows an additional Yukawa
coupling of the form $ f_LL^T\tau_2{\bf \tau}L.{\bf \Delta}$, where the
$\Delta^0$ couples to the neutrinos. Clearly $\Delta$ field has $L=2$. 
When $\Delta^0$ field has a
nonzero vev, it breaks lepton number by two units and leads to Majorana
mass for
the neutrinos. There are two questions that arise now: one, how does the
vev arise in a model and how does one understand the smallness of the
neutrino masses in this scheme. There are two answers to the first
question: One can maintain exact lepton number symmetry in the model and
generate the vev of the triplet field via the usual ``mexican hat''
potential. There are two problems with this case. This leads to the
triplet Majoron which has been ruled out by LEP data on Z-width.
Though it is now redundant it may be worth pointing out that in this model
smallness of the neutrino mass is not naturally understood.

There is however another way to generate the induced vev keeping a large
but positive mass ($M_{\Delta}$) for the triplet Higgs boson and allowing
for a lepton number violating coupling $M\Delta^* H H$. In this case,
minimization of the potential induces a vev for the $\Delta^0$ field when
 the doublet field acquires a vev: 
\begin{eqnarray}
v_T\equiv <\Delta^0>= \frac{M v^2_{wk}}{M^2_{\Delta}}
\end{eqnarray}
Since the mass of the $\Delta$ field is invariant under $SU(2)_L\times
U(1)_Y$, it can be very large connected perhaps with some new scale of
physics. If we assume that $M_{\Delta}\sim M\sim 10^{13}$ GeV or so, we
get $v_T\sim$ eV. Now in the Yukawa coupling $ f_LL^T\tau_2{\bf
\tau}L.{\bf \Delta}$, since the $\Delta^0$ couples to the neutrinos,
its vev
leads to a neutrino mass in the eV range or less depending on the value of
the Yukawa couplings\cite{many}. We will see later when we discuss the
seesaw models that unlike those models, the neutrino mass in this case is
not hierarchically dependent on the charged fermion masses. This model is
more in the spirit of models with minimal grand unification and can in
fact be implemented in
models\cite{book} such as those based on the SU(5) group where there is no
natural place for the right handed neutrino.

There are however two potential problems with this kind of models: the
first one is agan a naturalness question and the second a more detailed
cosmological one.

\begin{center}
{\it II.3A Naturalness of the Higgs triplet models}
\end{center}
To fit present neutrino data, one would like the neutrino mass scale to be
roughly in the eV range. The neutrino mass formula in the triplet Higgs
models is $m_\nu\simeq \frac{v^2_{wk}}{M_T}$. To get an eV scale mass
therefore requires that $M_T\simeq 10^{13}$ GeV. The qustion then arises
asto why scalar mass is of order of $10^{13}$ GeV rather than the Planck
scale. It is the same kind of question that leads to the so called gauge
hierarchy problem of the standard model. Since the Higgs triplet is a
scalar boson, one cannot use chiral symmetry arguments for this purpose.
One must necessarily make it part of a supersymmetric theory. This then
leads to a problem with baryogenesis that we discuss now.

\bigskip

 \begin{center}
{\it II.3B  Baryogenesis problem in models without right handed neutrinos}
\end{center}

  The simplest scenario
for baryogenesis in these models is via leptogenesis. The  only 
possibility\cite{sarkar} here is that
the decay of triplet Higgs to leptons provides a way to generate enough
baryons in the model. However for that to happen, one must satisfy one of
Sakharov's three conditions for matter-antimatter asymmetry i.e. the decay
particle which leads to baryon
or lepton asymmetry must be out of equilibrium. This requires that
\begin{eqnarray}
\frac{f^2_L M_{\Delta}}{12\pi}< \sqrt{g^*}\frac{M^2_{\Delta}}{M_{P\ell}}
\end{eqnarray}
implying a lower limit on the mass $M_{\Delta}\geq \frac{f^2_L
M_{P\ell}}{12\pi \sqrt{g^*}}$. For $f_L\sim 10^{-1}$ as would be required
by the atmospheric neutrino data, one gets conservatively, $M_{\Delta}\geq
10^{13}$ GeV. The problem with such a large mass arises from the fact
that in an inflationary model of the universe, the typical reheating
temperature dictated by the gravitino problem of supergravity is at most
$10^9$ GeV. Thus there is an inherent conflict between the standard
inflationary picture of the universe and the baryogenesis in the simple
triplet model for neutrino masses. For this model to work therefore, one
must invoke a new scenario to resolve the gravitino reheating problem. 

\begin{center}
{\bf II.4 A gauge model for Seesaw mechanism: left right symmetric
unification}
\end{center}
Let us now explore the implications of including the righthanded neutrinos
into the
extensions of the standard model to understand the small neutrino mass by
the seesaw mechanism. As already emphasized, if we assume that there are
no new symmetries
beyond the standard model, the right handed neutrino will have a natural
mass of order of the Planck scale making the light neutrino masses too
small to be of interest in understanding the observed oscillations. We
must therefore search for new symmetries that can keep the RH
neutrinos at a lower scale than the Planck scale. A new symmetry always
helps in making this natural.

To study this question, let us note that the inclusion of the right
handed neutrinos transforms the dynamics of the gauge models in a profound
way. To clarify what we mean, note that in the standard model (that does
not contain a $\nu_R$) the $B-L$ symmetry is only linearly anomaly free
i.e. $Tr[(B-L)Q^2_a]=0$ where $Q_a$ are the gauge generators of the
standard model but $Tr(B-L)^3\neq 0$. This means that $B-L$ is only a
global symmetry and cannot be gauged. However as soon as the $\nu_R$ is
added to the standard model, one gets $Tr[(B-L)^3]=0$ implying that the
B-L symmetry is now gaugeable and one could choose the gauge group of 
nature to be either $SU(2)_L\times U(1)_{I_{3R}}\times U(1)_{B-L}$ or
$SU(2)_L\times SU(2)_R\times U(1)_{B-L}$, the latter being the gauge group
of the left-right symmetric models\cite{moh}. Furthermore the presence of
the $\nu_R$ makes the model quark lepton symmetric and leads to a
Gell-Mann-Nishijima like formula for the elctric charges\cite{marshak}
i.e.
\begin{eqnarray}
Q= I_{3L}+I_{3R}+\frac{B-L}{2}
\end{eqnarray}
The advantage of this formula over the charge formula in the standard
model charge formula is that in this case all entries have a physical
meaning. Furthermore, it leads naturally to Majorana nature of neutrinos
as can be seen by looking at the distance scale where the $SU(2)_L\times
U(1)_Y$ symmetry is valid but the left-right gauge group is broken. In
that case, one gets
\begin{eqnarray}
\Delta Q=0= \Delta I_{3L}:\\ \nonumber
\Delta I_{3R}~=~-\Delta \frac{B-L}{2}
\end{eqnarray}
We see that if the Higgs fields that break the left-right gauge group
carry righthanded isospin of one, one must have $|\Delta L| = 2$ which
means that the neutrino mass must be Majorana type and the theory will
break lepton number by two units.

Let us now proceed to give a few details of the left-right symmetric model
and demonstrate how the seesaw mechanism emerges in this model.

The gauge group of the theory is
SU$(2)_L \, \times$ SU$(2)_R \, \times$ U$(1)_{B-L}$ with quarks and
leptons transforming as doublets under SU$(2)_{L,R}$.
In Table 3, we denote the quark, lepton and Higgs
fields in the theory along with their transformation properties
under the gauge group.
~~~~~~~~~~
\begin{center}
{\bf Table II}
\end{center}

\begin{center}
\begin{tabular}{|c|c|} \hline\hline
Fields           & SU$(2)_L \, \times$ SU$(2)_R \, \times$ U$(1)_{B-L}$ \\
                 & representation \\ \hline
$Q_L$                & (2,1,$+ {1 \over 3}$) \\
$Q_R$            & (1,2,$ {1 \over 3}$) \\
$L_L$                & (2,1,$- 1$) \\
$L_R$            & (1,2,$- 1$) \\
$\phi$     & (2,2,0) \\
$\Delta_L$         & (3,1,+ 2) \\
$\Delta_R$       & (1,3,+ 2) \\
\hline\hline
\end{tabular}
\end{center}

\noindent{\bf Table caption} Assignment of the fermion and Higgs
fields to the representation of the left-right symmetry group.

\bigskip

The first task is to specify how the left-right symmetry group breaks to
the standard model i.e. how one
breaks the $SU(2)_R\times U(1)_{B-L}$ symmetry so that the successes of
the standard model
including the observed predominant V-A structure of weak interactions at
low energies is reproduced. Another question of naturalness that also
arises simultaneously is that since the charged fermions and the
neutrinos are treated completely symmetrically (quark-lepton symmetry)
in this model, how does one understand the smallness of the neutrino
masses compared to the other fermion masses.

It turns out that both the above problems of the LR model have a common
solution. The process of spontaneous breaking of the $SU(2)_R$ symmetry
that suppresses the V+A
currents at low energies also solves the problem of ultralight neutrino
masses. To see this let us write the Higgs fields explicitly:
\begin{eqnarray}
\Delta~=~\left(\begin{array}{cc} \Delta^+/\sqrt{2} & \Delta^{++}\\
\Delta^0 & -\Delta^+/\sqrt{2} \end{array}\right); ~~
\phi~=~\left(\begin{array}{cc} \phi^0_1 & \phi^+_2\\
\phi^-_1 & \phi^0_2 \end{array}\right)
\end{eqnarray}
 All these
Higgs fields have Yukawa couplings to the fermions given symbolically as
below.
\begin{eqnarray}
{\cal L_Y}= h_1 \bar{L}_L\phi L_R +h_2\bar{L}_L\tilde{\phi}L_R\nonumber \\
+ h'_1\bar{Q}_L\phi Q_R +h_2'\bar{Q}_L\tilde{\phi}Q_R
\nonumber\\
+f(L_LL_L\Delta_L +L_RL_R\Delta_R) +~ h.c. \end{eqnarray}
The $SU(2)_R\times U(1)_{B-L}$ is broken down to the standard model
hypercharge $U(1)_Y$ by choosing $<\Delta^0_R>=v_R\neq 0$ since this
carries
both $SU(2)_R$ and $U(1)_{B-L}$ quantum numbers. It gives mass to the
charged and neutral righthanded gauge bosons i.e. $M_{W_R}= gv_R$ and
$M_{Z'}=\sqrt{2} gv_R cos\theta_W/\sqrt{cos 2\theta_W}$. Thus by
adjusting the value of $v_R$ one can suppress the right handed current
effects in both neutral and charged current interactions arbitrarily
leading to an effective near maximal left-handed form for the charged
current weak interactions.

The fact that at the same time the neutrino masses also become small can
be
seen by looking at the form of the Yukawa couplings. Note that the f-term
leads to a mass for the right handed neutrinos only at the scale $v_R$.
Next as we break the standard model symmetry by turning on the vev's for
the $\phi$ fields as $Diag<\phi>=(\kappa, \kappa')$, we not only
give masses to the $W_L$ and the $Z$ bosons but also to the quarks and the
leptons. In the neutrino sector the above Yukawa couplings after
$SU(2)_L$ breaking by $<\phi>\neq 0$ lead to the so called Dirac masses
for
the neutrino
connecting the left and right handed neutrinos. In the two component
neutrino language, this leads to the following mass matrix for the
$\nu, N$ (where we have denoted the left handed neutrino by $\nu$ and the
right handed component by $N$).
\begin{eqnarray}
M=\left(\begin{array}{cc} 0 & h\kappa \\
h\kappa & fv_R\end{array}\right)
\end{eqnarray}
Note that $m_D$ in previous discussions of the seesaw formula (see Eq. ())
is given by $m_D=h\kappa$, which links it to the weak scale and the mass
of the RH neutrinos is given by
$M_R=f v_R$, which is linked to the local B-L symmetry. This
justifies keeping RH neutrino mass at a scale lower than the Planck mass.
It is therefore fair to assume that seesaw mechanism coupled with
observations of neutrino oscillations are a strong indication of the
existence of a local B-L symmetry far below the Planck scale.

By diagonalizing this $2\times 2$ matrix, we get the light neutrino 
eigenvalue to be $m_{\nu}\simeq \frac{(h\kappa)^2}{fv_R}$ and the heavy
one to be $fv_R$. Note that typical
charged fermion masses are given by $h'\kappa$ etc. So since $v_R\gg
\kappa, \kappa'$, the light neutrino mass is automatically suppressed.
This way of suppressing the neutrino masses is called the seesaw mechanism
\cite{grsyms}. Thus in one stroke, one explains the smallness of the
neutrino mass as well as the suppression of the V+A currents.

In deriving the above seesaw formula for neutrino masses, it has been
assumed that the vev of the lefthanded triplet is zero so that the
$\nu_L\nu_L$ entry of the neutrino mass matrix is zero. However, in most
explicit models such as the left-right model which provide an explicit
derivation of this formula, there is an induced ve for the $\Delta^0_L$
of order $<\Delta^0_L> = v_T\simeq \frac{v^2_{wk}}{v_R}$. In the
left-right models, this this arises from the presence of a coupling in the
Higgs potential of the form
$\Delta_L\phi\Delta^{\dagger}_R\phi^{\dagger}$. In the  presence
of the $\Delta_L$ vev, the seesaw formula undergoes a fundamental
change. One can have two types of seesaw formulae depending on whether
the $\Delta_L$ has vev or not:

\bigskip
\noindent {\bf \underline{Type I seesaw formula}}
\bigskip

\begin{eqnarray}
M_{\nu}\simeq - M^T_{D}M^{-1}_{N_{R}}M_D
\end{eqnarray}
where $M_D$ is the Dirac neutrino mass matrix and $M_{N_R}\equiv fv_R$ is
the right handed neutrino mass matrix in terms of the $\Delta$ Yukawa
coupling matrix $f$.

\bigskip
 \noindent{\bf \underline{Type II seesaw formula}}
\bigskip

\begin{eqnarray}
M_{\nu}\simeq f\frac{v^2_{wk}}{v_R} - M^T_{D}M^{-1}_{N_{R}}M_D
\end{eqnarray}

Note that in the type I seesaw formula, what appears is the square of the
Dira neutrino mass matrix which in general expected to have the same
hierarchical structure as the corresponding charged fermion mass matrix.
In fact in some specific GUT models such as SO(10), $M_D=M_u$. This is the
origin of the common statement that neutrino masses given by the seesaw 
formula are hierarchical
i.e.
$m_{\nu_e}\ll m_{\nu_{\mu}}\ll m_{\nu_{\tau}}$ and even a more model
dependent statement that $m_{\nu_e} : m_{\nu_{\mu}} : m_{\nu_{\tau}}=
m^2_u : m^2_c : m^2_t$.

On the other hand if one uses the type II seesaw formula, there is no
reason to expect a hierarchy and in fact if the neutrino masses turn out
to be degenerate as discussed before as one possibility, one possible way
to understand this may be to use the type II seesaw formula.

Secondly, the type II seesaw formula is a reflection of the parity
invariance of the theory at high energies. Evidence for it would point
more strongly towards left-right symmetry at high energies.

\begin{center}
{\bf II.5~~  Understanding  detailed pattern for neutrinos using the
seesaw formula}
\end{center}

Let us now address the question: to what extent one can understand the
details of the neutrino masses and mixings using the seesaw formulae.
 The answer to this question is quite model dependent. While 
 there exist many models which fit the observations, none 
(except a few) are completely predictive and almost always they need
to invoke new symmetries or new assumtions. The problem in general is that  
the seesaw formula of type I, has 12
parameters in the absence of CP violation (six parameters for a symmetric
Dirac mass matrix and six for the $M_R$) which is why its predictive
power is so limited. In the presence of CP violation, the number
of parameters double making the situation worse. Specific predictions can
be made only under additional assumptions. 

For instance, 
in a class of seesaw models based on the SO(10) group that embodies the
left-right symmetric unification model or the SU(4)-color, the mass the
tau neutrino mass can be estimated provided one assumes the normal mass
hierarchy for neutrinos and a certain parameter accompanying a higher 
dimensional operator to be of order one. To see this, let us assume that
in the SO(10) theory, the B-L symmetry is broken by a {\bf 16}-dim. Higgs
boson. The RH neutrino mass in such a model arises from the
nonrenormalizable operator $\lambda{ \bf (16_F\bar{16}_H)^2/M_{P\ell}}$.
In a supersymmetric theory, if {\bf 16}-Higgs is also responsible for GUt
symmatry breaking, then after symmetry breaking, one obtains the RH
neutrino mass $M_R\simeq \lambda (2\times 10^{16})^2/M_{P\ell}\simeq
4\lambda 10^{14}$ GeV. In models with $SU(4)_c$ symmetry,
$m_{\nu_{\tau},D} \simeq m_t (M_U)\sim 100$ GeV. Using the seesaw formula
then, one obtains for $\lambda =1$, tau neutrino mass $m_{\nu_\tau}\simeq
.025$ eV, which is close to the presently preferred value of 0.05 eV. 
The situation with respect to other neutrino masses is however less
certain and here one has to make assumptions.

The situation with respect to mixing angles is much more
complicated. For instance, the striking difference between
the quark and neutrino mixing angles makes one doubt whether complete
quark lepton unification is truely obeyed in nature. In II.8 and III.3,
two examples are given where very few assumptions are made in getting 
maximal atmospheric mixing angle.

\begin{center}
{\bf II.6~~  General consequences of the seesaw formula for
neutrino masses}
\end{center}

In this section, we will consider some implications of the seesaw
mechanism for understanding neutrino masses. We will discuss two main
points. One is the nature of the right handed neutrino spectrum as 
dictated by the seesaw mechanism and secondly, ways to get an
approximate $L_e-L_\mu-L_\tau$ symmetric neutrino mass matrix using the
seesaw mechanism and its possible implications for physics beyond the
standard model\cite{falcone}. 

For this purpose, we use the type I seesaw formula along with the
assumption of a diagonal Dirac neutrino mass matrix to obtain the 
right handed neutrino mass matrix $M_R$:
\begin{eqnarray}
{\cal M}_{R, ij}~=~ m_{D,i}\mu^{-1}_{ij}m_{D,j}
\end{eqnarray}
with
\begin{eqnarray}
\mu^{-1}_{11} = \frac{c^2}{m_1}+\frac{s^2}{m_2}+\frac{\epsilon^2}{m_3} \\
\nonumber
\mu^{-1}_{12} = -\frac{c(s+ c\epsilon)}{\sqrt{2}m_1} +\frac{s(c-
s\epsilon)}{\sqrt{2}m_2} +\frac{\epsilon}{\sqrt{2}m_3}
\\ \nonumber
\mu^{-1}_{13} = \frac{c(s- c\epsilon)}{\sqrt{2}m_1} - \frac{s(c+
s\epsilon)}{\sqrt{2}m_2} +\frac{\epsilon}{\sqrt{2}m_3} \\ \nonumber
\mu^{-1}_{22} = \frac{(s+c\epsilon)^2}{{2}m_1} +\frac{(c_
s\epsilon)^2}{{2}m_2} + \frac{1}{2m_3} \\ \nonumber
\mu^{-1}_{23} =-\frac{(s^2- c^2\epsilon^2)}{{2}m_1}-\frac{(c^2-
s^2\epsilon^2)}{{2}m_2} +\frac{1}{2m_3} \\ \nonumber
\mu^{-1}_{33} = \frac{(s- c\epsilon)^2}{{2}m_1}+\frac{(c+
s\epsilon)^2}{{2}m_2} +\frac{1}{2m_3}.
\end{eqnarray}
Since for the cases of normal and inverted hierarchy, we have no
information on the mass of the lightest neutrino $m_1$, we could assume it
in principle to be quite small. In that case, the above equation enables
us to conclude that quite likely one of the three right handed neutrinos
is much heavier than the other two. The situation is of course completely
different for the degenerate case. This kind of sseparation of the RH
neutrino spectrum is very suggestive of a symmetry. In fact we have
recently argued that\cite{kuchi}, this indicates the possible existence of
an $SU(2)_H$ horizontal symmetry, that leads in the simplest case to an
inverted mass pattern for light neutrinos. This idea is discussed in a
subsequent section.

\begin{center}
{\bf II.7~~  $L_e-L_\mu-L_\tau$ symmetry and $3\times 2$ seesaw}
\end{center}

In this section, we discuss how an approximate $L_e-L_\mu-L_\tau$
symmetric neutrino mass matrix may arise within a seesaw framework.
Consider a simple extension of the standard model by adding two
additional singlet right handed neutrinos\cite{lavoura}, $N_1, N_2$
assigning
them $L_e-L_\mu-L_\tau$ quantum numbers of $+1$ and $-1$
respectively. Denoting the standard model lepton doublets by
$\psi_{e,\mu,\tau}$, the $L_e-L_\mu-L_\tau$ symmetry allows the
following new couplings to the Lagrangian of the standard model:
\begin{eqnarray}
{\cal L}'~=~(h_3\bar{\psi_{\tau}} + h_2\bar{\psi}_{\mu})H N_2 +
h_1\bar{\psi}_eHN_1 + M N^T_1C^{-1}N_2 + h. c.
\end{eqnarray}
where $H$ is the Higgs doublet of the standard model; $C^{-1}$ is the
Dirac charge conjugation matrix.
We add to it the symmetry breaking mass terms for the right handed
neutrinos, which are soft terms, i.e.
\begin{eqnarray}
{\cal L}_B~=~ \epsilon(M_1 N^T_1C^{-1}N_1 + M_2 N^T_2C^{-1}N_2) + h.c.
\end{eqnarray}
with $\epsilon \ll 1$. These terms break  $L_e-L_\mu-L_\tau$ by two units
but since they are dimension 3 terms, they are soft and do not induce any

with $\epsilon \ll 1$. These terms break  $L_e-L_\mu-L_\tau$ by two units
but since they are dimension 3 terms, they are soft and do not induce any
new terms into the theory.

It is clear from the resulting mass matrix for the $\nu_L, N$ system
that the linear combination $h_2\nu_{\tau}-h_3\nu_{\mu})$ is massless and
the
atmospheric oscillation angle is given by $tan \theta_A = h_2/h_3$; for
$h_3\sim h_2$, the $\theta_A$ is maximal. The seesaw mass matrix then
takes the following form (in the basis $(\nu_e, \tilde{\nu}_{\mu}, N_1,
N_2)$ with $\tilde{\nu}_{\mu}\equiv h_2 \nu_{\mu} + h_3\nu_{\tau}$):
\begin{eqnarray}
M~=~\left(\begin{array}{cccc} 0 & 0 & m_1 & 0\\0 & 0 & 0 & m_2\\m_1 & 0 &
\epsilon M_1 & M \\ 0 & m_2 & M & \epsilon M_2\end{array}\right)
\end{eqnarray}
The diagonalization of this mass matrix leads to the mass matrix of the
form discussed before.

\begin{center}
{ \bf II.8~~  SO(10) realization of the seesaw mechanism}
\end{center}

The most natural grand unified theory for the seesaw mechanism is the
SO(10) model, although as has been mentioned, the gross dissimilarity
between quark and lepton mixings makes additional assumptions necessary 
 to reconcile with the inherent quark-lepton unification in such models.
Nevertheless, the SO(10) models are so natural framework for neutrino
masses that some salient features may be instructive for any neutrino
model building.

The first interesting point to note about the SO(10) models is that 
 the {\bf 16}-dimensional spinor representation contains
all the fermions of each generation in the standard model plus the right
handed
neutrino. Thus the right handed neutrino is necessary for the seesaw
mechanism is automatic in these models.
Secondly, in order to break the B-L symmetry present in the SO(10) group,
one may use either the Higgs multiplets in {\bf 16} or {\bf 126}
dimensional rep. We will see that either of these representations can 
be used to implement the seesaw mechanism. To see this note that under the
left-right symmetric
group $SU(2)_L\times SU(2)_R\times SU(4)_c$, these fields
decompose as follows:
\begin{eqnarray}
{\bf 16}= (2, 1, 4)\oplus (1,2, 4^*)\\ \nonumber
{\bf 126}= (1, 1, 6)\oplus (3,1, 10)\oplus (1, 3, 10^*)\oplus
(2,2, 15)  
\end{eqnarray}
Note that in order to break the B-L symmetry it is the $(1,2,4^*)$ and
$(1,3,10^*)$ in the respective multiplets
 whose neutral elements need to pick up a large vev. Note however
that the ${\bf 16}_H$ does not have any renormalizable coupling with the
{\bf
16} spinors which contain the $\nu_R$ whereas there is a renormalizable
SO(10) invariant coupling of the form ${\bf 16}{\bf 16}{\bf \bar{126}}$.
for the second multiplet. Therefore if we decided to stay with the
renormalizable model, then we
would need a {\bf 126} dimensional representation to implement the seesaw
mechanism whereas if we used the {\bf 16} to break the B-L symmtery, we
would require nonrenormalizable couplings of the form ${\bf 16}^2 {\bf
\bar{16}}^2/M_{P\ell}$. This has important implications for the B-L scale.
In the former case, the B-L breaking scale is at an intermediate level
such as $\sim 10^{13}$ GeV or so whereas in the latter case, we can have
B-L scale coincide with the GUT scale of $2\times 10^{16}$ GeV as in the
typical SUSYGUT models\cite{tasi}.

In addition to having the right handed neutrino as part of the basic
fermion representation and the Higgs representations the SO(10) model
several other potential advantages for understanding neutrino masses. 
For example, if one uses only the {\bf 10} dimensional representation for
giving masses to the quarks and leptons, one has the up
quark mass matrix $M_u$ being equal to the Dirac mass matrix of the
neutrinos which goes into the seesaw formula. As a result, if we work in a
basis where the up quark masses are diagonal so that all CKM mixings come
from the down mass matrix, then the number of arbitrary parameters in the
seesaw formula goes dowm from 12 to 6. Thus even though one cannot
predict neutrino masses and mixings, the parameters of
the theory get fixed by their values as inputs. This may then be testable
thru its other predictions. The {\bf 10} only Higgs models have problems
of their own i.e. there are tree level mass relations in the down sector such
as $\frac{m_d}{m_s}=\frac{m_e}{m_{\mu}}$ which are renormalization group
invariant and are in disagreement with observations. It may be possible in
supersymmetric models to generate enough one loop corrections out of the
supersymmetry breaking terms (nonuniversal) to save the situation but they
will introduce unknown parameters into the theory.

There is one very special class of models where all 12 parameters of the
neutrino sector are predicted by the
quark and lepton masses\cite{babum}. This is the minimal renormalizable
SO(10) model with the Higgs content of 
only one {\bf 10} and one {\bf 126} plus two Higgs multiplets {\bf
45}+{\bf 54}. The last two multiplets break the SO(10) symmetry down
to $SU(2)_L\times SU(2)_R\times U(1)_{B-L}\times SU(3)_c$. The {\bf 126}
breaks the$SU(2)_L\times SU(2)_R\times U(1)_{B-L}\times SU(3)_c$ down to
the standard model which is then broken by the {\bf 10} Higgs. 

SO(10) has the property that the Yukawa couplings involving the {\bf 10}
and {\bf 126} Higgs representations are symmetric\cite{book1}. Therefore
if we ignore CP violation and work in a basis where one of these two sets
of Yukawa coupling matrices is diagonal, then it will have
only nine parameters. Noting the fact that the (2,2,15) submultiplet of
{\bf 126} has a standard model doublet that contributes to charged fermion
masses, one can write the quark and lepton mass matrices as follows:
\begin{eqnarray}
M_u~=~ h \kappa_u + f v_u \\  \nonumber
M_d~=~ h \kappa_d + f v_d \\  \nonumber
M_\ell~=~ h \kappa_d -3 f v_d \\  \nonumber
M_{\nu_D}~=~ h \kappa_u -3 f v_u \\  
\end{eqnarray}
where $h$ is the {\bf 10} Higgs Yukawa coupling matrix and $f$ is the {\bf
126} Yukawa coupling matrix; $\kappa_{u,d}$ are the vev's of the up and
down type Higgs doublets in the {\bf 10} Higgs and $v_{u,d}$ are the
corresponding vevs for the standard model doublets in the {\bf 126} Higgs
multiplet.
Note that there are 13 parameters in the above equations and there are 12
inputs (six quark masses, three lepton masses and three quark mixing
angles). Thus all parameters except one is determined. 

Next important observation on this model is that the same {\bf 126}
responsible for the fermion masses also has a vev along the $\nu_R\nu_R$
directions so that it generates the right handed neutrino mass matrix
which are proportional to the $f$ matrix. Thus using the seesaw formula
(type I), there are no free parameters in the light neutrino sector. This
model was extensively analysed in \cite{babum} prior to the emergence of
all the neutrino oscillation data in late 90's. It was shown that this
minimal SO(10) model without any CP phase cannot fit both the solar and
the atmospheric neutrino data simultaneously and is therefore ruled out.
It has however been recently noted that once the CP phases are properly
included in the discussion, this model can yield a near bimaximal mixing
pattern for neutrinos\cite{fukuyama}. A generic prediction of these class
of models is a large (close to experimental upper limit) value for
$U_{e3}$. Therefore more refined measurements for this parameter under
planning can test this model. 

A very interesting point regarding these models has recently been noted
in Ref.\cite{bajc}, where they point out that if one assume that the
direct triplet term in type II seesaw, dominates, then it provides a very
natural understanding of the large atmospheric mixing angle. The simple
way to see it is to note that when the triplet term dominates the seesaw
formula, then we have the neutrino mass matrix ${\cal M}_\nu \propto f$,
where $f$ matrix is the {\bf 126} coupling to fermions discussed earlier.
Using the above equations, one can derive the following
sumrule\cite{brahma}:
\begin{eqnarray}
{\cal M}_\nu~=~ c (M_d - M_\ell)
\end{eqnarray}
Now quark lepton symmetry implies that for the second and third
generation, the $M_{d,\ell}$ have the following general form:
\begin{eqnarray}
M_d~=~\pmatrix{\epsilon_1 & \epsilon_2\cr \epsilon_2 & m_b}
\end{eqnarray}
and
\begin{eqnarray}
M_\ell~=~\pmatrix{\epsilon'_1 & \epsilon'_2\cr \epsilon'_2 & m_\tau}
\end{eqnarray}
where $\epsilon_i \ll m_{b,\tau}$ as is required by low energy
observations. It is well known that in supersymmetric theories, when low
energy quark and lepton masses are extrapolated to the GUT scale, one gets
approximately that $m_b\simeq m_\tau$. One then sees from the above
sumrule for neutrino masses that all entries for the neutrino mass matrix
are of the same order leading very naturally to the atmospheric mixing
angle to be large.

Few cautionary words are in order: 

(i) In this approach, one must make sure that indeed in the seesaw formula
for neutrinos, the triplet term indeed dominates.

(ii) This appraoch does not lead to the atmospheric mixing angle to be
maximal as is observed but that is it large.

(iii) Finally, since in this model there are no free parameters, the solar
mixing angle should be predicted and one has to see whether it is
large. In fact one such attempt before did give a small mixing angle,
although there might be other domains of parameters where one may get a 
large solar mixing angle.

 Recently other SO(10) models have been
considered where under different assumptions, the atmospheric and solar
neutrino data can be explained together\cite{many1}.

In the minimal supersymmetric left-right model, an analogous situation 
happens where the neutrino Dirac masses are found to be equal to the
charged lepton masses\cite{dutta}. Thus in this model too, one has only
six parameters to describe the neutrino sector and once the neutrino data
is fitted all parameters in the model are fixed so that one has
predictions that can be tested. For instance, it has been emphasized in
Ref.\cite{dutta} that there is a prediction for the $B(\tau\rightarrow
\mu+\gamma)$ in this model that is about two orders of magnitude below the
present limits \cite{cleo} 
and could therefore ne used to test the model.

Finally let us comment that in models where the light neutrino mass is
understood via the seesaw mechanism that uses heavy righthanded neutrinos,
there is a very simple mechanism for the generation of baryon asymmetry of
the universe. Since the righthanded neutrino has a high mass, it decays at
a high temperature to generate a lepton asymmetry\cite{fuku} and this
lepton asymmetry is converted to baryon asymmetry via the sphaleron
effects\cite{kuzmin} at lower temperature. It also turns out that one of
the necessary conditions for sufficient leptogenesis is that the right
handed neutrinos must be heavy as is required by the seesaw mechanism. To
see this note that one of Sakharov conditions for leptogenesis is that the
right handed neutrino decay must be slower than the expansion rate of the
universe at the temperature $T\sim M_{N_R}$. The corresponding condition
is:
\begin{eqnarray}
\frac{h^2_{\ell}M_{N_R}}{16\pi}\leq \sqrt{g^*}\frac{M^2_{N_R}}{M_{P\ell}}
\end{eqnarray}
This implies that $M_{N_R}\geq \frac{h^2_{\ell}M_{P\ell}}{16\pi
\sqrt{g^*}}$. For the second generation, it implies that
$M_{N_{2R}}\geq 10^{13}$ GeV and for the third generation a value even
higher. In arriving at this conclusion, we have assumed that leptonic
Yukawa couplings are of the same order as the up quark sector. Note that
the deduced lower limits on the RH neutrino masses are above
the inflation reheating upper bound from considerations of the gravitino
production 
alluded to before. However for the first generation, it can be about
$10^{8}$
GeV so that there is no conflict with the gravitino bound on the reheating
temperature and therefore leptogenesis can occur. Incidentally, the
leptogenesis condition also imposes limits
on the matrix elements of the right handed neutrino mass, thereby reducing
the arbitrariness of the seesaw predictions slightly.

\begin{center}
{\bf II.9~~  Type II seesaw and Quasi-degenerate neutrinos}
\end{center}

In this subsection we like to discuss some issues related to the
degenerate
neutrino hypothesis, which will be necessary if there is evidence
for neutrinoless double beta decay at a significant level(see for
example the recent results from the Heidelberg-Moscow
group\cite{klap}) and assuming that no other physics such as
R-parity breaking or doubly charged Higgs etc are not the source
of this effect). Thus it is appropriate to discuss how such models can arise in
theoretical schemes and how stable they are under radiative corrections.

There are two aspects to this question: one is whether the degeneracy
arises within a gauge theory framework without arbitrary adjustment of
parameters and the second aspect being that given such a degeneracy arises
at some scale naturally in a field theory, is this mass degeneracy stable
under renormali9zation group extrapolation to the weak scale where we need
the degeneracy to be present. In this section we comment on the first
aspect.
 
It has already been alluded to before and first made in
\cite{cald1}
is that degenerate neutrinos arise naturally in models that employ the
type
II seesaw since the first term in the mass formula is not connected to the
charged fermion masses. One way that has been discussed is to consider
schemes where one uses symmetries such as SO(3) or SU(2) or permutation
symmetry $S_4$\cite{deg} so that the Majorana Yukawa couplings $f_i$ are
all equal. This then leads to the dominant contribution to all
neutrinos being equal. This symmetry however must be broken in the charged
fermion sector in order to explain the observed quark and lepton masses.
Such models consistent
with known data have been constructed based on SO(10) as well as other
groups. The interesting point about the SO(10) realization is that the
dominant contributions to the $\Delta m^2$'s in this model comes from the
second term in the type II seesaw formula which in simple models is
hierarchical. It is of course known that if the MSW solution to the solar
neutrino puzzle is the right solution (or an energy independent solution),
then we have $\Delta m^2_{solar} \ll \Delta m^2_{ATMOS}$. In fact if we
use the fact true in SO(10) models that $M_u=M_D$, then we have $\Delta
m^2_{ATMOS}\simeq m_0\frac{m^2_t}{fv_R}$ and $\Delta
m^2_{SOLAR}\simeq m_0\frac{m^2_c}{fv_R}$ where $m_0$ is the common mass
for the three neutrinos. It is interesting that for $m_0\sim $ few eV and
$fv_R\approx 10^{15}$ GeV, both the $\Delta m^2$'s are close to the
required values.

Outside the seesaw framework, there could also be electroweak symmetries
that guarantee the mass degeneracy. For a recent model of this type see,
ref\cite{ma}.

The second question of stability under RGE of such a pattern is discussed
in a subsequent section.

\newpage

\section{}

\begin{center}
{\bf III.1  Lepton flavor violation and neutrino masses}
\end{center}

In the standard model, the masslessness of the neutrino implies that
that there is no lepton flavor changing effects unlike in the quark
sector. Once one includes the right handed neutrinos $N_R$ one for each
family, there is lepton mixing and therefore lepton flavor changing
 effects such as $\mu\rightarrow e+\gamma$, $\tau\rightarrow e,\mu
+\gamma$ etc. However, a simple estimate of the one loop contribution to
such effects shows that the amplitude is of order
\begin{eqnarray}
A(\ell_j\rightarrow \ell_i +\gamma) \simeq \frac{eG_F m_{\ell_j}m_e
m^2_\nu}{\pi^2m^2_W} \mu_B
\end{eqnarray}
This leads to an unobservable branching ratio (of order $\sim 10^{-40}$) 
for the rare radiative decay modes for the leptons.

The situation however changes drastically as soon as the seesaw mechanism
for neutrino masses is embedded into the supersymmetric models. 
It has been noted in many papers already that in supersymmetric theories,
the lepton flavor changing effects get significantly enhanced.
They arise from the the mixings among sleptons (superpartners of leptons)
of different flavor caused by
the renormalization group extrapolations which via loop diagrams lead to
lepton flavor violating (LFV) effects at low energies\cite{lfv}.

 The way this happens is as follows. In the simplest N=1 supergravity
models\cite{nath}, the supersymmetry breaking terms at the Planck scale
are taken to have only few parameters: a universal scalar mass $m_0$,
universal $A$ terms, one gaugino mass $m_{1/2}$ for all three types of
gauginos. Clearly, a universal scalar mass implies that at Planck scale,
there is no flavor violation anywhere except in the Yukawa couplings (or
when the Yukawa terms are diagonalized, in the CKM
angles). However as we extrapolate this theory to the weak scale, the
flavor mixings in the Yukawa interactions induce non universal flavor
violating scalar mass terms (i.e. flavor violating slepton and squark mass
terms). In the absence of neutrino masses, the Yukawa matrices for leptons
can be diagonalized so that there is no flavor violation in the lepton
sector even after extrapolation down to the weak scale. On the other hand,
when neutrino mixings
are present, there is no basis where all leptonic flavor mixings can be
made to
disappear. In fact, in the most general case, of the three matrices
$Y_{\ell}$, the
charged lepton coupling matrix, $Y_{\nu}$, RH neutrino Yukawa coupling and
$M_{N_R}$, the matrix characterizing the heavy RH neutrino mixing, only
one can
be diagonalizd by an appropriate choice of basis and the flavor mixing in
the other two remain. In a somewhat restricted case where the right
handed neutrinos do not have any interaction other than the Yukawa
interaction and an interaction that generates the Majorana mass for
the right handed neutrino, one can only diagonalize two out of the
three matrices (i.e. $Y_\nu, Y_\ell$ and $M_R$). Thus there will always 
be lepton flavor violating terms in the basic Lagrangian, no matter
what basis one chooses. These LFV terms can then induce mixings
between
the sleptons of different flavor and lead to LFV processes. If we keep the
$M_{\ell}$ diagonal by choice of basis,
searches for LFV processes such as $\tau\rightarrow \mu +\gamma$ and/or
$\mu\rightarrow e +\gamma$ can throw light on the RH neutrino mixings/or
family mixings in $M_D$, as has already been observed.

Since in the absence of CP violation, there are at least six mixing angles
(nine if $M_D$ is not symmetric) in the seesaw formula and only three are
observable in neutrino oscillation, to get useful information on the
fundamental high scale theory from LFV processes, it is assumed that
$M_{N_R}$ is diagonal so that one has a direct correlation between
the observed neutrino mixings and the fundamental high scale paramters of
the theory. The important point is that the flavor mixings in $Y_{\nu}$
then reflect themselves in the slepton mixings that lead to the LFV
processes via the RGEs. 

From the point of view of the LFV analysis, there are essentially two
classes of neutrino mass models that need to be considered: (i) the first
class is where
it is assumed that the RH neutrino mass $M_{N_R}$ is either a mass term in
the basic Lagrangian or arises from nonrenormalizable terms such as
${\nu^c\chi^c}^2/M_{P\ell}$, as in a class of SO(10) models; and (ii) a
second class where the Majorana mass of the right handed neutrino itself
arises from a renormalizable Yukawa coupling e.g. $f\nu^c\nu^c
\Delta$. In the first class of models, in principle, one could decide to
have all the flavor mixing effects in the right handed neutrino mass
matrix and keep the $Y_\nu$ diagonal. In that case, RGEs would not induce
any LFV effects. However we will bar this possibility and consider the
case where all flavor mixings are in the $Y_\nu$ so that RGEs can induce
LFV effects and estimate them in what follows. In class two models on the
other hand, there will always be an LFV effect, although its magnitude
will depend on the choice of the seesaw scale ($v_{BL}$).

Examples of class two models are models for neutrino mixings
 such as SO(10) with a {\bf 126} Higgs field\cite{babum} or
left-right
model with a triplet Higgs, whose vev is the seesaw scale.

In both these examples, the key equations that determine the extent of
lepton flavor violation are:

\noindent{\it case (i):}
\begin{eqnarray}
\frac{dm^2_L}{dt}~=~\frac{1}{4\pi^2}[ (m^2_L +
2m^2_{H_d}) Y^{\dagger}_\ell Y_\ell 
+(m^2_L+2H^2_u)Y^{\dagger}_{\nu}Y_{\nu} +2Y^{\dagger}_\ell m^2_{e^c}Y_\ell
+Y^{\dagger}_\ell Y_\ell m^2_\ell\\ \nonumber +2Y^{\dagger}_\nu
m^2_{\nu^c}Y_\nu
+Y^{\dagger}_\nu Y_\nu m^2_\ell
+2A^{\dagger}_\ell A_\ell +2A^{\dagger}_\nu A_\nu - f(g^2)]\\ \nonumber
\frac{dA_\ell}{dt}~=~ \frac{1}{16\pi^2} A_\ell [ Tr (3Y^{\dagger}_dY_d +
Y^{\dagger}_\ell Y_\ell) + 5 Y^{\dagger}_\ell Y_\ell+Y^{\dagger}_\nu Y_\nu
-3g^2_2-a g^2_{R}-b g^2_{B-L}]\\ \nonumber 
+Y_\ell[Tr(6A_dY^{\dagger}_d+A_\ell
Y^{\dagger}_\ell) + 4 Y^{\dagger}_\ell A_\ell+ 2Y^{\dagger}_\nu
A_\nu+ 6g^2_2M_2 +...]
\end{eqnarray}

\noindent{\it Case (ii)}
In addition to the above two equations, two more equations are necessary:
\begin{eqnarray}
\frac{dY_{\nu}}{dt}~=~\frac{Y_{\nu}}{16\pi^2}
[Tr(3Y_uY^{\dagger}_u+Y_{\nu}Y^{\dagger}_{\nu}) +3Y^{\dagger}_{\nu}Y_{\nu}
 +Y^{\dagger}_{\ell}Y_{\ell}+4f^{\dagger}f -3g^2_2-c_R
g^2_{R}-c_{B-L}g^2_{B-L}] \\\nonumber
\frac{dA_\nu}{dt}~=~ \frac{1}{16\pi^2} A_\nu [ Tr (3Y^{\dagger}_uY_u +
Y^{\dagger}_\nu Y_\nu) + 5 Y^{\dagger}_\nu Y_\nu+Y^{\dagger}_\ell Y_\ell
+ 4f^{\dagger}f-3g^2_2-a g^2_{R}-b g^2_{B-L}]\\ \nonumber
+Y_\ell[Tr(6A_dY^{\dagger}_d+A_\ell
Y^{\dagger}_\ell) + 4 Y^{\dagger}_\ell A_\ell+6g^2_2M_2 +...)
\end{eqnarray}
In order to apply these equations, we note that in the basis where the
charged lepton masses are diagonal, the seesaw formula involves the right
handed neutrino mass matrix $M_R$ and the neutrino Dirac mass matrix
$M_D$. Assuming the the Dirac mass matrix is symmetric, there are 12
parameters( for the case with CP conservation). Since neutrino masses and
mixings only provide six observables, there are several different ways
that can lead to the observed neutrino mixings. Two distinct extreme 
ways are as follows: (i) the first case is where
the neutrino mixings arise primarily from the off diagonal elements
of the $M_D$ assuming the $M_R$ is diagonal or even an extremely
simplified case where it is a unit matrix and (ii) a second case where
we can keep the $M_D$ diagonal and all mixings arise from $M_R$ having
mixings. In the first case, the RGEs always lead to lepton flavor
violation whereas in the second case, flavor violations arise only if the
$M_R$ arises from a Majorana Yukawa coupling of the form
$\nu^c\nu^c\Delta$ after $<\Delta>\neq 0$ as already explained. We will
call this the Majorana case and case (i) as the Dirac case.

In the Dirac case, starting with the simplest supersymmetry breaking
assumption of universal scalar masses and proportional $A$ terms, the 
scalar sleptons develop off diagonal terms due to the flavor violation in
$Y_\nu$ and these mass terms have the form
\begin{eqnarray}
m^2_{\tilde{L},ij}\propto \frac{3+a^2}{16\pi^2}\ell
n\frac{M_{P\ell}}{M_{B-L}}Y^{\dagger}_\nu Y_\nu + \cdot\cdot\cdot
\end{eqnarray}
where $\cdot\cdot\cdot$ denote the diagonal terms that cannot cause flavor
mixing.
If the $M_R$ is diagonal, the $Y^{\dagger}_\nu Y_\nu$ is nothing but the
neutrino mass matrix upto a constant $=M_R$
Using this we can get the Dirac mass dependence of the
$B(\ell_j\rightarrow e + \gamma)$ (for $\ell_j = \tau, \mu$) to be:
\begin{eqnarray}
B(\ell_j\rightarrow e + \gamma) \sim \frac{(m_2cs+m_3 \epsilon)^2
v^2_{B-L}}{G^2_Fv^4_{wk}m^4_0} c_{\ell_j}
\end{eqnarray}
where $c_{\tau}=1/6$ and $c_{\mu}=1$ whereas for $B(\tau\rightarrow \mu +
\gamma ) \propto 6\frac{m^2_3 v^2_{B-L}}{4G^2_Fv^4_{wk}m^4_0}$.

Now coming to the second case with Majorana-Yukawa couplings,
starting with universal scalar masses at the Planck scale and in the
Majorana-Yukawa case all couplings flavor diagonal except the $f$
coupling, it is
easy to see that the above equations will induce flavor changing effects
in $m^2_{\tilde{L}}$ and $A_\ell$. The strength of the slepton flavor
mixings depends sensitively on the neutrino mass texture and the resulting
the texture in the coupling matrix $f$. Below we give the branching ratio
expressions for one extreme case where we assume that the lightest
neutrino mass dominates the $f$ matrix elements. We caution the reader
that this is by no means the most typical case and results differ
significantly as different mass textures are considered\cite{bdm6}.

The slepton flavor mixings, via a one loop diagram
involving the winos ($\tilde{W}^{+}$ and $\tilde{W}^3$ lead to the lepton
flavor changing radiative amplitudes. Keeping only the contribution of the
$m^2_{\tilde{L}}$ term which dominates for larger $tan\beta$,
we find that roughly speaking the three branching ratios are given by:
\begin{eqnarray}
B(\mu\rightarrow e + \gamma) \propto \frac{1}{G^2_Fm^4_0
v^4_{B-L}}\frac{m^4_{D3}m^6_{D2}m^2_{D1}tan^2\beta}{m^4_1v^4_{wk}};
\end{eqnarray}
\begin{eqnarray}
B(\tau\rightarrow e + \gamma) \propto \frac{1}{G^2_Fm^4_0
v^4_{B-L}}\frac{m^{10}_{D3}m^2_{D1}tan^2\beta}{m^4_1v^4_{wk}};
\end{eqnarray}
\begin{eqnarray}
B(\tau\rightarrow \mu + \gamma) \propto \frac{1}{G^2_Fm^4_0
v^4_{B-L}}\frac{m^{10}_{D3}m^2_{D2}tan^2\beta}{m^4_1v^4_{wk}}.
\end{eqnarray}
Note an important difference between the predictions for the Dirac case
and the Majorana one. Both $\tau\rightarrow e+\gamma$ and $\mu\rightarrow
e+\gamma$ are of the same order of magnitude for the Dirac case whereas
they are very different for the Majorana case where $B(\mu\rightarrow
e+\gamma)/B(\tau\rightarrow e+\gamma) \simeq
(m_{D2}/m_{D3})^6$. Therefore,
even a mild hierarchy in the Dirac mass sector can make the
$\tau\rightarrow e+\gamma$ branching ratio much larger than the
$\mu\rightarrow e+\gamma$ branching ratio. This kind of discrepancy could
in principle be used to test the origin of the seesaw mechanism.

It must be pointed out that the predictions for the Majorana case are
extremely texture sensitive. In a recent paper, \cite{bdm6}, calculations
have been carried out for a different but consistent texture where
$B(\mu\rightarrow e+\gamma)$ though much lower than the $\tau$ rare decay
branching ratios, is still found to be within the range of currently
planned experiments at PSI.

For completeness, we give the formula for calculating the radiative decay
of the leptons. If we express the amplitude for the decay as:
\begin{eqnarray}
{\cal L}~=~iem_j\left(\bar{\ell}_{jL}\sigma_{\mu\nu}\ell_{iR} C_L
+\bar{\ell}_{jR}\sigma_{\mu\nu}\ell_{iL}C_R\right) F^{\mu\nu} + h.c.
\end{eqnarray}
then the Branching ratio for the decay $\ell_j\rightarrow \ell_i+\gamma$
is given by the formula
\begin{eqnarray}
B(\ell_j\rightarrow
\ell_i+\gamma) ~=~\frac{48\pi^3\alpha_{em}}{G^2_F}(|C_L|^2+|C_R|^2)
B(\ell_j\rightarrow \ell_i+ 2\nu)
\end{eqnarray}

 \begin{center}
{\bf III.2  Renormalization group evolution of the neutrino mass
matrix}
\end{center}

In the seesaw models for neutrino masses, the neutrino mass arises from
the effective operator
\begin{eqnarray}
{\cal O}_\nu~=~-\frac{1}{4}\kappa_{\alpha\beta} \frac{L_\alpha HL_\beta
H}{M}
\end{eqnarray}
after symmetry breaking $<H^0>\neq 0$; here $L$ and $H$ are the leptonic
and weak doublets respectively. $\alpha$ and $\beta$ denote the weak
flavor index. The matrix $\kappa$ becomes the neutrino mass matrix
after symmetry breaking i.e. $<H^0>\neq 0$. This operator is defined at
the scale
$M$ since it arises after the heavy field $N_R$ is integrated out. On the
other hand, in conventional oscillation experiments, the neutrino masses
and mixings being probed are at the weak scale. One must therefore
extrapolate the operator down from the seesaw scale $M$ to the weak scale
$M_Z$\cite{blp}. The form of the renormalization group extrapolation of
course depends on the details of the theory. For simplicity we will
consider only the supersymmetric theories, where the only contributions
come from the wave function renormalization and is therefore easy to
calculate. The equation governing the extrapolation of the
$\kappa_{\alpha\beta}$ matrix is given in the case of MSSM by:
\begin{eqnarray} 
\frac{d\kappa}{dt} =[-3g^2_2+6Tr(Y^{\dagger}_uY_u)]\kappa
+\frac{1}{2}[\kappa(Y^{\dagger}_eY_e)+(Y^{\dagger}_eY_e)\kappa]
\end{eqnarray}
We note two kinds of effects on the neutrino mass matrix from the above
formula: (i) one that is flavor independent and (ii) a part that is flavor
specific. If we work in a basis where the charged leptons are diagonal,
then the resulting correction to the neutrino mass matrox is given by:
\begin{eqnarray}
{\cal M}_\nu (M_Z)~=~ (1 + \delta){\cal M}(M_{B-L}) (1 + \delta)
\end{eqnarray}
where $\delta$ is a diagonal matrix with matrix elements
$\delta_{\alpha\alpha} \simeq -\frac{m^2_{\alpha}tan^2\beta}{16\pi^2v^2}$
In more complicated theories, the corrections will be different. Let us
now study some implications of this corrections. For this first note that
in the MSSM, this effect can be sizable if $tan\beta$ is large (of order
10 or bigger).

\begin{center}
{\bf III.3  Radiative magnification of neutrino mixing angles}
\end{center}

A major puzzle of quark lepton physics is the diverse nature of the
mixing
angles. Whereas in the quark sector the mixing angles are small, for the
neutrinos they are large. One possible suggestion in this connection is
that perhaps the mixing angles in both quark and lepton sectors at similar
at some high scale; but due to renormalization effects, they may become
magnified at low scales. It was shown in ref.\cite{balaji} that this
indeed happens if the neutrino spectrum os degenerate. This can be seen in
a simple way for the $\nu_\mu-\nu_\tau$ sector\cite{balaji}.

Let us start with the mass matrix in the flavor basis:
\begin{eqnarray}
\cal{M_F}  =  U^* {\cal{M_D}} U^{\dag} \nonumber \\
   =  \left(\begin{array}{cc} C_\theta & S_\theta \\ -S_\theta &
C_\theta \end{array} \right)
       \left(\begin{array}{cc} m_1 & 0 \\ 0 & m_2 e^{-i\phi} \end{array}
\right)
        \left(\begin{array}{cc} C_\theta & -S_\theta \\ S_\theta &
C_\theta \end{array} \right).
\label{uudag}
\end{eqnarray}
 Let us examine the situation when $\phi=0$ (i.e. CP is conserved), which
corresponds to the
case when the neutrinos $\nu_1$ and $\nu_2$ are in
the same $CP$ eigenstate.
Due to the presence of radiative corrections to $m_1$ and $m_2$,
the matrix $\cal{M_F}$ gets modified to
\begin{eqnarray}
\cal{M_F} \to \left(\begin{array}{cc} 1+\delta_\alpha & 0 \\ 0 &
1+\delta_\beta \end{array} \right) \cal{M_F}
     \left(\begin{array}{cc} 1+\delta_\alpha & 0 \\ 0 & 1+\delta_\beta
\end{array} \right).
\label{mf-fin}
\end{eqnarray}
 The mixing angle $\bar\theta$ that now diagonalizes the matrix
$\cal{M_F}$ at the low scale $\mu$ (after radiative corrections)
can be related to the old mixing angle $\theta$
through the following expression:
\begin{eqnarray}
\tan 2\bar\theta~=~\tan 2\theta~(1~+~\delta_\alpha + \delta_\beta)
~\frac{1}{\lambda}~,
\label{t2tbar}
\end{eqnarray}
where
\begin{eqnarray}
\lambda \equiv
\frac{(m_2~-~m_1)C_{2\theta}~+~2\delta_\beta(m_1S_\theta^2~+~
m_2C_\theta^2)~-~2\delta_\alpha(m_1C_\theta^2~+~
m_2S_\theta^2)}{(m_2~-~m_1)C_{2\theta}}~.
\label{lam}
\end{eqnarray} 

If
\begin{eqnarray}
(m_1 - m_2)~ C_{2\theta}~=~2\delta_\beta(m_1S_\theta^2~+~
m_2C_\theta^2)~-~2\delta_\alpha(m_1C_\theta^2~+~
m_2S_\theta^2)~,
\label{cond}
\end{eqnarray}
then $\lambda = 0$ or equivalently
$\bar\theta = \pi/4$; {\it i.e.} maximal mixing.
Given the mass heirarchy of the charged leptons: $m_{l_\alpha} \ll
m_{l_\beta}$, we expect $|\delta_\alpha| \ll |\delta_\beta|$,
which reduces (\ref{cond}) to a simpler form:
\begin{eqnarray}
\epsilon~=~\frac{\delta m C_{2\theta}}{(m_1S_\theta^2~+~ m_2C^2_{\theta}}
\end{eqnarray}
In the case of MSSM, the radiative magnification condition can be
satisfied provided provided
\begin{eqnarray}
h_\tau (MSSM) ~\approx~ \sqrt{\frac{8 \pi^2|\Delta m^2(\Lambda)|
C_{2\theta}}
{ln(\frac{\Lambda}{\mu})m^2}}~.
\label{hbeta}
\end{eqnarray}
For $\Delta m^2|simeq \Delta m^2_A$, this ccondition can be satisfied for
a very wide range of $tan\beta$. 

It is important to emphasize that this magnification occurs only if at the
seesaw scale the neutrino masses are nearly degenerate. A similar
mechanism using the righthanded neutrino Yukawa couplings instead of the 
charged lepton ones has been carried out recently\cite{lindner}. Here two
conditions must be satisfied: (i) the neutrino psectrum must be nearly
degenerate (i.e. $m_1 \simeq m_2$ as in ref.\cite{balaji}) and (ii) there
must be a hierarchy between the righhanded neutrinos.

\begin{center}
{\bf III.4 An explicit example of a neutrino mass matrix unstable under
RGE}
\end{center}

In this section, we give an explicit example of a neutrino mass matrix
unstable under RGE effects\cite{lola}. Consider the following mass matrix
with degenerate neutrino masses and a bimaximal mixing\cite{georgi}.
\begin{eqnarray}
{\cal M}_\nu~=~\left(\begin{array}{ccc} 0 & \frac{1}{\sqrt{2}}
&\frac{1}{\sqrt{2}} \\ \frac{1}{\sqrt{2}} & \frac{1}{2} & -\frac{1}{2}\\
\frac{1}{\sqrt{2}} & \frac{1}{2} & \frac{1}{2} \end{array}\right)
\end{eqnarray}
The eigenvalues of this mass matrix are $(1, -1, 1)$ and the eigenvectors:
\begin{eqnarray}
V_1=\left(\begin{array}{c} 0 \\ \frac{1}{\sqrt{2}}
\\\frac{1}{\sqrt{2}}\end{array}\right); 
V_2 = \left(\begin{array}{c}  \frac{1}{\sqrt{2}} \\ -\frac{1}{2} 
\\-\frac{1}{{2}}\end{array}\right);
V_3 = \left(\begin{array}{c} \frac{1}{\sqrt{2}}
\\\frac{1}{{2}} \\ \frac{1}{2}\end{array}\right)
\end{eqnarray}
After RGE to the weak scale, the mass matrix becomes
\begin{eqnarray}
{\cal M}_\nu~=~\left(\begin{array}{ccc} 0 & \frac{1}{\sqrt{2}}
&\frac{1}{\sqrt{2}}(1+\delta) \\ \frac{1}{\sqrt{2}} & \frac{1}{2} &
-\frac{1}{2}(1+\delta)\\
\frac{1}{\sqrt{2}}(1+\delta) & \frac{1}{2}(1+\delta) &
\frac{1}{2}(1+2\delta) \end{array}\right)
\end{eqnarray}
It turns out that the eigenvectors of this matrix become totally different
and are given by:
\begin{eqnarray}
V_1=\left(\begin{array}{c}  \frac{1}{\sqrt{3}}
\\\frac{2}{\sqrt{3}}\\ 0\end{array}\right);
V_2 = \left(\begin{array}{c}  \frac{1}{\sqrt{2}} \\ -\frac{1}{2}
\\-\frac{1}{{2}}\end{array}\right);
V_3 = \left(\begin{array}{c} \frac{1}{\sqrt{6}}
\\-\frac{1}{{2\sqrt{3}}} \\ \frac{\sqrt{3}}{2}\end{array}\right)
\end{eqnarray}
We thus see that the neutrino mixing pattern has become totally altered.
although the eigenvalues are only slightly perturbed from their
unperturbed value.

There are also other examples where the RGE of the neutrino mass matrix
can totally destabilize the mass pattern, see \cite{ibara} for an example
with the sterile neutrino.

\begin{center}
{\bf III.5  Generation of solar mass difference square in the
case of $L_e-L_\mu-L_\tau$ symmetric models}
\end{center}

In this section, we give another example where the renormalization group
evolution (RGE) of the neutrino mass matrix plays an important role.
Consider the following neutrino mass matrix defined at a (unification)
scale much above the weak scale as in Eq. (6).
\begin{eqnarray}
{\cal M}_\nu^0 = \left(\matrix{0 & 1 & 1 \cr 1 & 0 & 0 \cr 1 & 0 &
0}\right) m~.
\end{eqnarray}
Note that as already emphasized earlier, this matrix has several deficits
as far as accomodating neutrino data is concerned. To remind the reader,
the problems are two fold: (i) it leads to $\Delta m^2_{\odot}=0$ and (ii)
solar neutrino mixing angle is maximal. As we show below, RGEs can help to
overcome both these defects.

The important point to note is that the RGE depends on the charged lepton
mass matrix. Let us assume that the charged lepton mass matrix 
has the form:
\begin{eqnarray}
{\cal M}_{\ell^+} = \left(\matrix{ 0 & 0 & x \cr 0 & y & 0 \cr x' & 0 &
1}\right) m_\tau~.
\end{eqnarray}
For the most part we will take $x'=x$ so that $|x| \simeq
\sqrt{m_e/m_\tau}$
and $y \simeq m_\mu/m_\tau$.  We will also consider the possibility that
$x \gg x'$ (the right handed singlet leptons multiply the matrix in
this equation on the right). In the case $x=x'$, note that there is only
parameter in the
leptonic sector (both the charged leptons and neutrinos). We will show
that this model can lead to a realistic description of the neutrino
oscillations.

At scales below the unification scale, through the renormalization of
the effective $d=5$ neutrino mass operator the form of
$M_\nu^0$\cite{babu} will be modified.
(Analogous corrections in $M_{\ell^+}$ is negligible.)
The modified neutrino mass matrix at the weak scale is given by

\begin{eqnarray}
{\cal M}_\nu \simeq {\cal M}_\nu^0 + {c \over 16 \pi^2} {\rm
ln}(M_U/M_Z)\left(
Y_\ell Y_\ell^\dagger M_\nu^0 + M_\nu^0 (Y_\ell Y_\ell^\dagger)^T\right)~.
\end{eqnarray}
Here $c=-3/2$ for SM while $c=1$ for SUSY, and $Y_{\ell}$ is the
charged lepton Yukawa coupling matrix.
We have absorbed the flavor--independent renormalization factor into
the definition of $m$ in the above equation.  Explicitly,
\begin{eqnarray}
{\cal M}_\nu \simeq \left(\matrix{2 r x & 1+r(x^2+y^2) & 1+r(1+2x^2) \cr
1+r(x^2+y^2) & 0 & rx \cr
1+r(1+2x^2) & rx & 2rx}\right)m ~,
\end{eqnarray}
This mass matrix has both an acceptable $\Delta m^2_{\odot}$ as well as
large (but not maximal) solar neutrino mixing angle. This another example
where the renormalization group equations do make a difference in our
understanding of the neutrino oscillations.

\begin{center} 
{\bf III.6  A horizontal symmetry approach to near bimaximal
mixing}
\end{center}

In this section, I present a model\cite{kuchi} which motivates the
existence of an $SU(2)_H$ horizontal symmetry acting on leptons to
understand the near bimaximal mixing patter and yields the
softly broken $L_e-L_\mu-L_\tau$ model discussed earlier. 

Suppose, there is an $SU(2)_H$ horizontal symmetry that acts only on
leptons. As already discussed, freedom from global Witten anomaly requires
that there must be two right handed neutrinos that trasform as a doublet
of $SU(2)_H$. The local $SU(2)_H$ symmetry then implies that the masses of
those two right handed neutrinos are protected and must be at the
scale of $SU(2)_H$ breaking. If there is a the third right handed neutrino
for reasons of quark lepton symmetry, then it will acquire a mass of order
of the Planck or string scale and decouple from neutrino physics at
lower energies. This
therefore provides a physically distinct way of implementing the seesaw
mechanism. One has a $3\times 2$ seesaw rather than the usual $3\times 3$
one.

Furthermore the $SU(2)$ horizontal symmetry restricts both the Dirac mass
of the neutrino as well as the righthanded neutrino mass matrix to the
forms\cite{kuchi}

\begin{eqnarray}
M_{\nu_L,\nu_R}~=~\left(\begin{array}{ccccc} 0 & 0 & 0 & h_0\kappa_0 & 0\\
0 & 0 & 0 & 0 & h_0\kappa_0\\ 0 & 0 & 0 & h_1\kappa_1 & h_1 \kappa_2 \\
h_0\kappa_0 & 0 & h_1\kappa_1 & 0 & fv'_H \\ 0 & h_0\kappa_0 & h_1\kappa_2
& fv'_H & 0 \end{array} \right)
\end{eqnarray}
After seesaw diagonalization, it leads to the light neutrino mass matrix
of the form:
\begin{eqnarray}
{\cal M}_{\nu}~=~-M_D M^{-1}_R M^T_D
\end{eqnarray}
where $M_D~=~\left(\begin{array}{cc} h_0\kappa_0 & 0 \\ 0 & h_0\kappa_0\\
h_1\kappa_1 & h_1\kappa_2
\end{array}\right)$; $M^{-1}_R~=~\frac{1}{fv'_H}\left(\begin{array}{cc} 0
&
1\\1 & 0 \end{array}\right)$. The resulting light Majorana neutrino mass
matrix ${\cal M}_{\nu}$ is given by:
\begin{eqnarray}
{\cal M}_{\nu}~=~-\frac{1}{fv'_H}\left(\begin{array}{ccc} 0 &
(h_0\kappa_0)^2 & h_0h_1\kappa_0\kappa_2\\ (h_0\kappa_0)^2 & 0 &
h_0h_1\kappa_0\kappa_1 \\ h_0h_1\kappa_0\kappa_2 & h_0h_1 \kappa_0\kappa_1
& 2h^2_1\kappa_1\kappa_2 \end{array}\right)
\end{eqnarray}
To get the physical neutrino mixings, we also need the charged lepton mass
matrix defined by $\bar{\psi}_L {\cal M}_\ell \psi_R$. This is given in
our model by:
\begin{eqnarray}
{\cal M}_{\ell}~=~\left(\begin{array}{ccc} h'_2\kappa_0 & 0 &-h'_1\kappa_2
\\ 0 & h'_2\kappa_0 & h'_1\kappa_1 \\ h'_4\kappa_1 & h'_4\kappa_2 &
h'_3\kappa_0 \end{array}\right)
\end{eqnarray}
Note that in the limit of $\kappa_1 =0$, the neutrino mass matrix
has exact $(L_e-L_{\mu}-L_{\tau})$ symmetry whereas the
charged lepton mass matrix breaks this symmetry. This is precisely the
class of inverted hierarchy models that was discussed earlier which
provides a realistic as well as a testable model for neutrino
oscillations. In particular, this model
leads to a relation between the neutrino parameters $U_{e3}$
and the ratio of solar and atmospheric mass difference squared i.e.
\begin{eqnarray}
U^2_{e3} cos 2\theta_{\odot} = \frac{\Delta m^2_{\odot}}{ 2 \Delta m^2_A}
+ O(U^4_{e3}, m_e/m_\mu)
\end{eqnarray}
which is testable in proposed long baseline
experiments such as NUMI off-axis plan at Fermilab or JHF in Japan.

\newpage

\section{}

\begin{center}
{\bf IV.1  Neutrino masses in models with large extra
dimensions}
\end{center}

One of the important predictions of string theories is the existence of
more than three space dimensions. For a long time, it was believed that
these extra dimensions are small and are therefore practically
inconsequential as far as low energy physics is concerned. However, recent
progress in the understanding of the nonperturbative
aspects of string theories have opened up the possibility that some of
these extra dimensions could be large\cite{horava,nima} without
contradicting observations. In particular, models where some of the extra
dimensions have sizes as large as a millimeter and where the
string scale is  in the few TeV range have attracted a great
deal of phenomenological attention in the past two years\cite{nima}. The
basic assumption of these models,
inspired by the D-branes in string theories, is that the
space-time has a brane-bulk
structure, where the brane is the familiar (3+1)
dimensional space-time, with the standard model particles and forces
residing in it, and the bulk consists of all space dimensions where
gravity and other possible gauge singlet particles live. One could of
course envision (3+d+1) dimensional D-branes where d-space dimensions have
miniscule ($\leq TeV^{-1}$) size.
The main interest in these models has been
due to the fact that the low string scale provides an opportunity
to test them using existing collider facilities.

A major challenge to these theories comes from the neutrino sector, the
first problem being how one understands the small
neutrino masses in a natural manner. The conventional
seesaw\cite{grsyms}
explanation which is believed to provide the most satisfactory way to
understand this, requires that the new physics scale (or the scale of
$SU(2)_R\times U(1)_{B-L}$ symmetry) be around $10^{9}$ to $10^{12}$ GeV
or higher, depending on the Dirac masses of the neutrinos whose magnitudes
are not known. If the highest scale of the theory is a TeV,
clearly the seesaw mechanism does not work, so one must look for
alternatives. The second problem is that if one considers only the
standard model group in the brane, operators such as $LH LH/M_*$ could be
induced by string theory in the low energy effective Lagrangian. For
TeV scale strings this would obviously lead to unacceptable neutrino
masses.

One mechanism suggested in Ref.\cite{dienes} is to postulate
the existence of one or more gauge singlet neutrinos, $\nu_B$, in the
bulk which couple to the lepton doublets in the brane. After
electroweak symmetry breaking, this coupling can lead to neutrino Dirac
masses, which are suppressed by the ratio $M_*/M_{P\ell}$, where
$M_{P\ell}$
is the Planck mass and $M_*$ is the string scale. This is
sufficient to explain small neutrino masses and owes its origin
to the large bulk volume that suppresses the effective Yukawa couplings of
the Kaluza-Klein (KK) modes of the bulk neutrino to the brane fields.
In this class of models, naturalness of small neutrino mass requires that
one must assume the existence of a global B-L symmetry in the theory,
since that will exclude the undesirable higher dimensional operators from
the theory.

To discuss the mechanisms in a concrete setting,
let us first focus on TeV scale models. Here, one postulates a bulk
neutrino, which is a singlet under the electroweak gauge group.
Let us denote the bulk neutrino by $\nu_B( x^{\mu}, y)$.  The bulk
neutrino
is represented by a four-component spinor and can be split into two chiral
Weyl 2-component spinors as $\nu^T_B = (\chi^T,
-i\phi^{\dagger}\sigma_2)$. The 2-component spinors $\chi$ and $\phi$ can
be decomposed in terms of 4-dimensional Fourier components as follows:
\begin{eqnarray}
\chi(x,y) = \frac{1}{\sqrt{2R}}\chi_{+,0} +\frac{1}{\sqrt{R}}
\sum_{n=1}^\infty
\left(\chi_{+,n} cos \frac{n\pi y}{R} +i \chi_{-,n} sin \frac{n\pi
y}{R}\right).
\end{eqnarray}
There is a similar expression for $\phi$.
It has a five dimensional kinetic energy term and a coupling to the brane
field $L(x^{\mu})$.  The full Lagrangian involving the $\nu_B$ is
\begin{eqnarray}
 {\cal L} = i\bar{\nu}_B\gamma_{\mu}\partial^{\mu}\nu_B+ \kappa  \bar{L} H
\nu_{BR}(x, y=0) +i \int dy\
 \bar{\nu}_{BL}(x,y)\partial_5 \nu_{BR}(x,y) + h.c.,
 \label{l1}
 \end{eqnarray}
where $H$ denotes the Higgs doublet, and
$\kappa = h {M_*\over M_{P\ell}}$ is the suppressed Yukawa coupling. This
leads to a Dirac mass for the neutrino\cite{dienes} given by:
\begin{eqnarray}
m = \frac{h v_{wk} M_*}{M_{P\ell}},
\label{mvsmstar}\end{eqnarray}
where $v_{wk} $ is the scale of $SU(2)_L$ breaking.
In terms of the 2-component fields, the mass term coming from the fifth
component of the kinetic energy connects the fields $\chi_+$ with $\phi_-$
and $\chi_-$ with $\phi_+$, whereas it is only the $\phi_+$ (or
$\nu_{B,R,+}$) which couples to the brane neutrino $\nu_{e,L}$. Thus as
far as the standard model particles and forces go, the fields $\phi_-$ and
$\chi_+$ are totally decoupled, and we will not consider them here. The
mass matrix that we will write below therefore connects only $\nu_{eL}$,
$\phi_{+,n}$ and $\chi_{-,n}$.

From Eq. \ref{mvsmstar}, we conclude that for $M_*\sim 10$ TeV, this leads
to
$m \simeq 10^{-4} h$ eV. It is
encouraging that this number is in the right range to be of interest in
the discussion of solar neutrino oscillation if the Yukawa coupling $h$
is appropriately chosen. Furthermore, this neutrino is mixed with all the
KK modes of the bulk neutrino, with a mixing mass $\sim \sqrt{2} m$; since
the nth KK mode has a mass $nR^{-1}\equiv n\mu$, the mixing angle is given
by $\sqrt{2} mR/n$. Note that for $R\sim 0.1 mm$, this mixing angle is of
the right order to be important in MSW transitions of solar neutrinos.

It is worth pointing out that this suppression of $m$
is independent of the number and radius
hierarchy of the extra dimensions, provided that our bulk neutrino
propagates in the whole bulk. For simplicity, we will
assume that there is only one extra dimension with radius of
order of a millimeter.

Secondly, the above discussion can be extended in a very straight forward
manner to the case of three generations. The simplest thing to do is to
add three bulk neutrinos and consider the Lagrangia to be:
\begin{eqnarray}
 {\cal L} = i\bar{\nu}_{B,\alpha}\gamma_{\mu}\partial^{\mu}\nu_{B,\alpha}+
\kappa_{\alpha\beta} \bar{L}_{\alpha} H
\nu_{B\beta R}(x, y=0) +i \int dy\
 \bar{\nu}_{B\alpha L}(x,y)\partial_5 \nu_{B\alpha R}(x,y) + h.c.,
 \end{eqnarray}
One can now diagonalize $\kappa_{\alpha\beta}$ by rotating both the bulk
and the active neutrinos. The mixing matrix then becomes the neutrino 
mixing matrix ${\bf \Large U}$ discussed in the text. In this basis (the
mass eigenstate basis), one can diagonalize the mass matrix
involving the $\nu_i$'s and the bulk neutrinos to get the mixings between
the active and the bulk tower. There are now three mixing parameters, one
for each mass eigenstate denoted by $\xi_i\equiv \sqrt{2} m,_i R$ and
mixing angle for each mass eigenstate to the nth KK mode of the
corresponding bulk neutrinos is given by $\xi/n$. 
The obseved oscillation data can then be used to put limits on $\xi_i$;
we discuss this in a subsequent section.

It is also worth noting that due to the presence of the infinite tower
mixed with the active neutrino, the oscillation probabilities are
distorted in a way which is very different from the case of oscillation
to a single neutrino level. This has the implication that if at some
point the complete oscillation of a neutrino (as opposed to just the
overall suppression as is the case now) is observed, it will be possible
to put stronger limits on the parameter $\xi_i$ from data.

\noindent{\bf IV. 1A: Neutrino propagation in matter with a bulk neutrino
tower}

Let us discuss the propagation of a neutrino in matter in the bulk tower
neutrino models. For this purpose we have to consider the neutrino mass
matrix in the flavor basis in the presence of matter effect, to be denoted
by $\delta_{ee}$. This looks as follows in the basis:  $(\nu_e,
\nu^{(0)}_{BR,+},
\nu^{(1)}_{BL,-}, \nu^{(1)}_{BR,+}, \nu^{(2)}_{BL,-}, \nu^{(2)}_{BR,+},
\cdot
\cdot)$ is given by:
\begin{eqnarray}
{\cal M}~=~ {\cal M}_{TeV} ~\equiv~
\left(\begin{array}{cccccccc} \delta_{ee} & m & 0 & \sqrt{2}m & 0 &
\sqrt{2}m & \cdot & \cdot\\
m & 0 & 0 & 0 & 0 & 0 & \cdot & \cdot \\
0 & 0 & 0 & \mu & 0 & 0 & \cdot & \cdot \\
\sqrt{2}m & 0 & \mu & 0 & 0 & 0 & \cdot & \cdot \\
0 & 0 & 0 & 0 & 0 & 2\mu & \cdot & \cdot \\
\sqrt{2}m & 0 & 0 & 0 & 2\mu & 0 & \cdot & \cdot \\
\cdot & \cdot & \cdot & \cdot & \cdot & \cdot & \cdot &\cdot \\
\cdot & \cdot & \cdot & \cdot & \cdot & \cdot & \cdot &\cdot \end{array}
\right). \label{MTeV}
\end{eqnarray}
where $\delta_{ee}$ is a possible matter effect.
One can evaluate the eigenvalues and the eigenstates of this matrix. The
former are the solutions of the transcendental equation:
\begin{eqnarray}
m_n~=~\delta_{ee} + \frac{\pi m^2}{\mu}cot\left(\frac{\pi
m_n}{\mu}\right).
\label{mnTeV}
\end{eqnarray}
The equation for eigenstates is
\begin{eqnarray}
\tilde \nu_n =\frac{1}{N_n}\left[ \nu_e + {m\over m_n}\nu^{(0)}_{B,+}
+
\sum_k\sqrt{2}m \left(\frac{m_n}{m^2_n - k^2\mu^2}\nu^{(k)}_{B,-} +
\frac{k\mu}{m^2_n-k^2\mu^2} \nu^{(k)}_{B,+}\right)\right],
\label{nus}
\end{eqnarray}
where we have used the notation $\pm$ for the left- and right-handed parts
of the KK modes of the bulk neutrino in the two-component notation and
dropped the $L,R$ subscripts,
the sum over $k$ runs through the KK modes, and
$N_n$ is the normalization factor given by
\begin{eqnarray}
N_n^2 = 1 + m^2\pi^2R^2 + {(m_n-\delta_{ee})^2\over m^2}.
\label{Nn}
\end{eqnarray}

One of the very striking effects of the KK tower of bulk neutrinos is the
presence of multiple MSW resonances each time the neutrino energy is such
that one satisfies the resonance condition in a medium. As in the two
neutrino matter resonance, each time there is a level crossing by the
$\nu_e$ due to its matter effect, there will be a dip in the survival
probability of the electron neutrino.

 To understand
the origin of the dips, note that the MSW resonance condition is given by
\begin{eqnarray}
m^2_{res} \simeq \frac{4G_F\rho}{\sqrt{2} m_p} E_{res}
\end{eqnarray}
As a result, as the neutrino energy $E$ increases, the resonance
condition is satisfied for higher and higher KK modes of the bulk
neutrino. The survival probability at and after the resonance is given by
\begin{eqnarray}
P_{ee} \simeq e^{-\frac{\pi m^2_n sin^2 2 \theta}{2E} R_{eff}}
\end{eqnarray}
for $E\geq E_{res}$ and $P_{ee}\sim 1$ for $E< E_{res}$. The cumulative
survival probability is given by $P=
P^{(1)}P^{(2)}\cdot\cdot\cdot P^{(n)}$. Note that due the exponent
being
larger than one at the resonance (which can be checked by putting
numbers), as soon as a new KK level is crossed, the $P_{ee}$ dips to
a value much less than one and then starts to rise as $E$ increases giving
rise to the dip structure. The net of effect of the combined dip structure
is to flatten the spectrum.

 The typical values of the survival probability within the
$^8B$ region ($\sim 6$ to $\sim 14$ MeV) are quite sensitive to the value
of
$mR$.  As can be seen from Eq. \ref{Nn}, higher $mR$
increases $1/N_n\approx m/m_n\approx mR/n$ for various
$n$, and thereby increases $\nu_e$ coupling to higher mass eigenstates,
strengthens MSW resonances, and lowers $\nu_e$ survival probability.
Thus searching for dips in solar neutrino spectrum is one way to probe the 
size of extra dimensions. The same dip phenomena could also appear in
higher energy atmospheric spectra. Since the resonance condition given
by the above formula implies that the higher the energy the higher the
mass difference squared (or lower the extra dimension radius) probed,
search for dips in higher energy neutrinos can in principle reveal the
existence of extra dimensions of smaller sizes that cannot be probed by
solar neutrinos.

\begin{center}
{\bf IV.2  Phenomenological and cosmological constraints on
bulk neutrino  models}
\end{center}

A generic feature of understanding neutrino mass via bulk neutrinos in
models with large flat extra dimension is the presence of an infinite
tower of closely spaced sterile neutrinos mixed with active neutrino (say
$\nu_e$). Information about the nature of extra dimensions can therefore
be obtained by looking at the phenomenological as well as cosmological
effects of this mixing with the infinite tower.

\subsection{Neutrino oscillation constraints}

To see the kind of constraints one can derive, let us take a simple model
where there are three bulk neutrinos (i.e. three infinite towers), each
giving mass to one family of neutrinos. In this minimal model, a simple
rotation of the active neutrinos takes the neutrinos to the mass basis
and separates three towers. It has been shown in recent
papers through detailed analysis\cite{apl}, that while this minimal model
can adequately explain both the solar and
atmospheric neutrino oscillation phenomena essentially by arranging the
active bulk neutrino mixing, it is not possible to accomodate the LSND results.
Furthermore, since generic mixing of the various
mass eigenstate neutrinos to the nth KK mode of the corresponding bulk
neutrino goes like $m_iR/n$, if $R$ is sizable, the observable effect
simulating a multitude of sterile neutrinos should be present. Since both
present atmospheric as well as solar neutral current data from SNO
severely constrain the oscillation of the active neutrinos to the sterile 
ones, this puts a constraint on $R$. For instance, considering the muon
neutrino oscillation to bulk neutrinos in the atmospheric neutrino data,
we can take $m_3 R \leq 10\%$. This implies that $R\leq 2$ eV$^{-1}$ or
$R$ less than 40 microns. This constraint is more stringent than the one
derived from direct searches for deviations from the inverse square
law\cite{edel}.

\subsection{Big Bang Nucleosynthesis constraints}

On the cosmological front, since big bang nucleosynthesis is a very
sensitive measure of the number extra neutrino species mixed with the
active neutrinos, one should be able to get information about this class
of models from this data. The basic idea here is that at the epoch of
nucleosynthesis, neutrino oscillation to the bulk modes can bring in the
modes into thermodynamic equilibrium with all other relativistic
species. If that happens each mode will contribute an amount $\rho_\nu$
to the energy density of the universe and speed up the expansion. The
faster the expansion of the universe, the earlier the weak interactions go
out of equilibrium (to be called freeze-out). Since the neutron to proton
ration depends very sensitively on the temperature of freeze-out
i.e. $n/p\sim e^{-\frac{m_n-m_p}{T_*}}$, the higher the freezeout
temperature $T_*$, the higher the neutron fraction and hence the Helium
fraction of the universe. 

Higher the mixing strength of the active modes with the bulk modes, the
more modes from the bulk that get into thermal equilibrium with the
electron. Therefore, to be consistent with observed Helium abundance, one
must have a restriction on the mixing parameter $\sqrt{2}mR$.

Very careful analysis of this has been done in
two papers\cite{goh} by solving the Boltzman equation for the
generation of sterile neutrinos from active neutrinos at the BBN
epoch. For the case of one space dimension, it was concluded in \cite{goh}
that the active-bulk mixing and the inverse radius of the extra dimension
$\mu$ satisfy the constraint:
\begin{eqnarray}
\left(\frac{\mu}{eV}\right)^{0.92}~ sin^22\theta  \leq 7.06 \times
10^{-4}
\end{eqnarray}
Applying this to the tau neutrino, we find an even more stringent limit
on the size of the extra dimensions $R$ than the one just listed
i.e. $R\leq 1.5$ micron. 

\subsection{Enhancement of magnetic moments} 

Another interesting consequence of the presence of the bulk tower is in
its effect on the magnetic moment of the active neutrinos. As is wellknown
\cite{shrock}, if one adds a singlet right handed neutrino to the standard
model to give a Dirac mass to the neutrino, this induces a magnetic moment
$\mu_{\nu}= 10^{-19}\frac{m_nu}{eV} \mu_B$, where $\mu_B$ is the Bohr
magneton ($\mu_B=e/2m_e c$). Since in the bulk neutrino models, there is
an infinite tower of neutrinos mixed with the active neutrinos, there is a 
$\mu_{\nu_e\nu_{B,p}}= 10^{-19}\frac{\sqrt{2}m}{eV}\mu_B$ connecting the
active neutrino with each KK mode of the bulk neutrino. In a neutrino
scattering process $\nu_e + e $ with neutrino energy $E_\nu$, all bulk
neutrino modes upto $E_\nu$ will be excited\cite{mcng}. Thus the effective
neutrino magnetic moment will appear to be $\mu_{eff}\simeq
10^{-19}\frac{m}{eV} (ER)^{1/2}$ where $R$ is the radius of the extra
dimension. For $R\sim$ millimeters, this enhances the magnetic moment by
almost a factor of a million. Thus this could ptovide an interesting way
to probe the existence of extra dimensions.  A note of caution is that
this apparent enhancement is effective only when the KK modes of the bulk
neutrino can be excited. For instance, when neutrino spin precesses in a
magnetic field, the most of the KK modes do not get excited due to energy
momentum conservation and therefore, the magnetic moment enhancement 
does not take place.

\subsection{Other implications}
Another interesting implication of the bulk neutrino tower appears if the 
brane model is not the standard model with one Higgs doublet but with two.
In this case, the physical charged Higgs can decay into charged lepton anf
the bulk tower but with the difference that unlike in the normal two Higgs
extension, the final state charged lepton emitted in this process will
have left-handed helicity whereas in models without any bulk neutrino, the
final state charged lepton will have right handed helicity\cite{agashe}.

The presence of the bulk neutrino tower also leads to new contributions to
flavor changing leptonic rare decays such as $\mu\rightarrow e+\gamma$
etc. They in turn lead to constraints on the fundamental scale of
nature\cite{pila}.

 \begin{center}
{\bf IV.3  Neutrino mass in low scale gravity models without
bulk neutrinos}
\end{center}

Since the presence of light tower of bulk neutrinos is so
constraining and adhoc forbidding of fully allowed operators is 
theoretically unappealing for the brane
bulk picture with the standard model in the brane, it is important to
search for alternative ways to solve the neutrino mass problem. In
these kind of scenarios, the strategy is to search for higher
dimensional models that will lead to the standard model after
compactification of the extra dimensions and yet allow for the possibility
of a low fundamental scale. One such alternative has recently been
proposed 
by having the left-right symmetric model or an extension of the standard 
model with only a local B-L symmetry in the bulk and looking for the
standard model in the zero mode part of the spectrum\cite{perez}.
It is assumed that all fermions are in the bulk. The number of extra space
dimensions can either be one or two. The model with two extra space
dimensions has additional symmetry which also helps top solve the proton
decay problem. So we first give the example of the six dimensional
model with the gauge group $SU(2)_L\times U(1)_{I_{3R}}\times U(1)_{B-L}$. 
Such models are have been called universal extra dimension
models\cite{ponton}. 

The minimal fermion content of this model is dictated by gravitational
anomaly cancellation to be \cite{ponton}:

$$Q_{+}(2,0,1/3),~ \psi_{+}(2,0,-1),~
 U_{-}(1,1/2,1/3),~D_{-}(1,-1/2,1/3)$$,
$$~E_{-}(1,-1/2,-1),
~N_{-}(1,+1/2,-1)$$ 

where $Q=(u,d)$ and $\psi=(\nu, e)$ and $\pm$ denote
the six dimensional chirality;
the numbers in the parentheses are the gauge quantum numbers.
Note that each fermion field is a four component field with two
4-dimensional 2
component spinors with opposite chirality e.g. $Q$ has a left chiral $Q_L$
and a right chiral field $Q_R$. As such the theory is
vectorlike at this stage and we will need orbifold projections to obtain a
chiral theory. We choose one Higgs doublet $\phi(2,-1/2,0)$ and a singlet
$B-L$ carrying Higgs boson $\chi(1,1/2,-1)$.  We compactify the theory on
a $T_2/Z_2$ orbifold; where $T_2$ is defined by the extra
coordinates $y_{1,2}$ satisfying the following conditions: $y_{1,2}
= ~y_{1,2}+2\pi R$ and $Z_2$ operates on the two extra coordinates
 as follows: $(y_1,y_2)\rightarrow (-y_1,-y_2)$. We now impose the
orbifold conditions on the fields as follows: We choose the following
fields to be even under the $Z_2$ symmetry: $Q_L,\psi_L, U_R, D_R, E_R,
N_L$; the kinetic energy terms then force the opposite chirality states
to be odd under $Z_2$. Note specifically that, in contrast with the
$U,D,E$ fields, it is the $N_L$ which is chosen even under $Z_2$. This is
crucial to our understanding of neutrino masses. As is well known, the
even fields when Fourier expanded involve only the $cos \frac{ny}{R}$ and
the odd fields only $sin \frac{ny}{R}$. As a result only the $Z_2$ even
fields  have zero modes. Thus, with the above compactification, below the
mass scale $R^{-1}$, the only fermionic modes are those of the standard
model plus the $N^{0}_L$.
When we give vev to the field $<\chi>= ~v_{BL}$, it
breaks the group down to the standard model. We will choose $v_{B-L}\sim
800$ GeV to a TeV.

Before discussing the implications of the model for neutrino masses and
proton decay, let us study the extra symmetries of the 4-dimensional
theory implied by the fact that it derives from a 6-dimensional one.

First, the discrete translational symmetry insures the conservation
of the fifth and sixth momentum components, $p_a$,
which are  quantized in integer factors of $1/R$.

Secondly, in the full uncompactified six dimensional theory, there is an
extra $U(1)_{45}$ symmetrry associated with the rotations in the $y_1,y_2$
plane. After compactification, the $U(1)_{45}$ invariance reduces to a
$Z_4$ symmetry. Therefore invariance under the $SO(1,3)\times Z_4$
space-time Lorentz transformations must be imposed on all possible
operators allowed in the effective four dimensional theory i.e. the
allowed operators will be those that  are  invariant under the whole
$SO(1,5)$ symmetry,
plus probably those for which the sum of fermion $U(1)_{45}$ charges
is equal to zero modulus 8.
The reasoning is as follows:
the $Z_4$  spatial symmetry,
actually translates into a $Z_8$ symmetry group
for the spinorial representation. In fact under a
$\pi/2$ rotation of the
$x_4$-$x_5$ plane a fermion
transforms as $\Psi(x') = U\Psi(x)$;
with $U= \exp[i(\pi/2)\Sigma_{45}/2]$;
where  $\Sigma_{45}= i[\Gamma^4,\Gamma^5]/2$
is the generator of the $U(1)_{45}$ group.

To see which operators are allowed, we need to know the $U(1)_{45}$
quantum numbers of the theory which can be easily read of from the six
dimensional theory and are given in the table below.

\bigskip
\begin{center}

{\bf Table III}
\end{center}

\begin{center}

\begin{tabular}{|c||c|}\hline
$Q_L,\psi_L$ & +1/2\\ \hline
$U_R,D_R,E_R,N_R$ & +1/2 \\ \hline
$Q_R,\psi_R, U_L$ & -1/2 \\ \hline
$E_L, D_L, N_L$ & -1/2 \\ \hline\hline

\end{tabular}

\end{center}

{\bf Table caption}: $U(1)_{45}$ charges of the various fermions 
in the  $SU(2)_L\times U(1)_{I_{3R}}\times U(1)_{B-L}$ model.

We will use these quantum numbers below.

Let us now turn to understanding the small neutrino mass in this
model. Note
that in this model due to our orbifold assignments and choice of the
gauge group coupled with the residual $Z_8$ symmetry discussed above, we
only have one term that to leading order can contribute to neutrino masses
and the term is:$\lambda \frac{\psi^T_L C^{-1}N_L\phi
(\chi^*)^2}{M^5_*}$ .

The following potentially dangerous terms are forbidden for various
reasons in this 6-dimensional theory:
\begin{itemize}

\item $(\psi_L\phi)^2/M_*$  is forbidden by $B-L$ symmetry.

\item Terms like $\bar{\psi}_L\phi N_R$,
though allowed are also not problematic
due to $Z_2$ quantum numbers which imply that the $N_R$ has no zero modes.

\item $\frac{(\psi_L\phi)^2(\chi^*)^2}{M^7_*}$ and $N_LN_L(\chi^*)^2$ are
forbidden by the residual $Z_8$ symmetry.

\end{itemize}

These operators are written in the 6-dimensional field theory. Upon
compactification to the 4-dimensional theory on our orbifold, the
 operator that leads to neutrino mass has the form  $\lambda
\frac{\psi^{(0)T}_L C^{-1}N^{(0)}_L\phi^{(0)}
(\chi^{(0)*})^2}{M^5_*R^3}$ .
Using $M_*\simeq 100$ TeV and $R^{-1}\sim $ TeV and using $\lambda \sim
0.1$, we find for that it leads to $m_\nu\sim $ eV, which is in the right
range without any fine tuning. Furthermore, the neutrinos in this model
are Dirac particles since all Majorana terms are forbidden to leading
order by the $Z_8$ symmetry.

\subsection{Left-right symmetric model in five dimensions and neutrino
mass}
In this section, we provide a left-right symmetric embedding of the local
B-L symmetry and show in a five dimensional example how it can lead to
small neutrino masses despite the fact that the fundamental scale of the
model is in the TeV range. These considerations are easily extended to the
six space-time dimensions.

Two basic new ingredients of the model are : $SU(2)_L\times SU(2)_R \times
U(1)_{B-L}$ gauge groiup and a slightly different orbifold
compactification based on $S_1/(Z_2\times Z'_2)$ for the five dimensional
model. In this model,
 one needs two sets of 4-component spinors for each family of
quarks and leptons. A requirement is that the left-right gauge symmetry is
broken by orbifold compactification\cite{nandi}. For this purpose, one
compactifies the 5th dimension on an $S_1/Z_2\times Z'_2$. It is well
known that for this compactification, the Fourier modes of the five
dimensional fields can be labelled as $(+,+), (+,-), (-,+), (-,-)$ where
$+$ and $-$ denote the $Z_2\times Z'_2$ parity of a given mode. They are
associated with $cos\frac{2ny}{R}, cos\frac{(2n-1)y}{R},
sin\frac{(2n-1)y}{R}$ and $sin\frac{2ny}{R}$ respectively. It is then easy
to see that only the $(+,+)$ models have zero mass and all other modes
have masses proportional to $R^{-1}$.
The various fermion fields are assigned the following $Z_2\times Z'_2$
quantum numbers:
\begin{eqnarray}
 Q_{1,L}\equiv
   \left(\begin{array}{c} u_{1L}(+,+)\\ d_{1L}(+,+)\end{array}\right);
 &\quad&
 Q'_{1,L}\equiv
   \left(\begin{array}{c} u'_{1L}(+,-)\\ d'_{1L}(+,-)\end{array}\right);
   \nonumber \\
 Q_{1,R}\equiv
   \left(\begin{array}{c} u_{1R}(-,-)\\ d_{1R}(-,-)\end{array}\right);
 &\quad&
 Q'_{1,R}\equiv
   \left(\begin{array}{c} u'_{1R}(-,+)\\ d'_{1R}(-,+)\end{array}\right);
   \nonumber \\ [1ex]
 Q_{2,L}\equiv
   \left(\begin{array}{c} u_{2L}(-,-)\\ d_{2L}(-,+)\end{array}\right);
 &\quad&
 Q'_{2,L}\equiv
   \left(\begin{array}{c} u'_{2L}(-,+)\\ d'_{2L}(-,-)\end{array}\right);
   \nonumber \\
 Q_{2,R}\equiv
   \left(\begin{array}{c} u_{2R}(+,+)\\ d_{2R}(+,-)\end{array}\right);
& \quad&
Q'_{2,R}\equiv
   \left(\begin{array}{c} u'_{2R}(+,-)\\ d'_{2R}(+,+)\end{array}\right);
\label{quarks}
\end{eqnarray}
and for leptons:
 \begin{eqnarray}
 \psi_{1,L}\equiv
   \left(\begin{array}{c} \nu_{1L}(+,+)  \\ e_{1L}(+,+)\end{array}\right);
  &\qquad&
 \psi'_{1,L}\equiv
   \left(\begin{array}{c} \nu'_{1L}(-,+)  \\
e'_{1L}(-,+)\end{array}\right);
   \nonumber \\
 \psi_{1,R}\equiv
    \left(\begin{array}{c} \nu_{1R}(-,-)  \\
e_{1R}(-,-)\end{array}\right);
  & \qquad &
 \psi'_{1,R}\equiv
  \left(\begin{array}{c} \nu'_{1R}(+,-) \\ e'_{1R}(+,-)\end{array}\right);
   \qquad
   \nonumber \\ [1ex]
 \psi_{2,L}\equiv
   \left(\begin{array}{c} \nu_{2L}(-,+) \\ e_{2L}(-,-)\end{array}\right);
  &\qquad&
 \psi'_{2,L}\equiv
    \left(\begin{array}{c} \nu'_{2L}(+,+)\\
e'_{2L}(+,-)\end{array}\right);
  \nonumber \\
 \psi_{2,R}\equiv
    \left(\begin{array}{c} \nu_{2R}(+,-)\\ e_{2R}(+,+)\end{array}\right);
&   \qquad &
 \psi'_{2,R}\equiv
  \left(\begin{array}{c} \nu'_{2R}(-,-)\\ e'_{2R}(-,+)\end{array}\right).
\label{leptons}
\end{eqnarray}
We see from the above equation that 
the process of compactification leaves us for $E\ll R^{-1}$,
 with the standard model fermions plus a neutrino like zero
mode that has only right handed weak
interaction, corresonding to a right handed doublet field dubbed as
$\psi'_2$. As far as the gauge bosons go, they have the following
``parity'' assignments:
\begin{eqnarray}
&&W_{1,\mu}^{3,\pm}(+,+);\quad B_\mu(+,+);\quad W^3_{2,\mu}(+,+);
 \quad W^\pm_{2,\mu}(+,-); \nonumber\\
&&W_{1,5}^{3,\pm}(-,-);\quad B_5(-,-);\quad W^3_{2,5}(-,-);
 \quad W^\pm_{2,5}(-,+).
\label{gparity}
\end{eqnarray}
Finally, the Higgs fields i.e. bidoublet field $\phi(2,2,0)$ and the
unidoublet fields $\chi_{L,R}$ have the following $Z_2\times Z'_2$
parities:
\begin{eqnarray}
\phi \equiv
\left(\begin{array}{cc} \phi^0_u(+,+) & \phi^+_d(+,-)\\
   \phi^-_u(+,+) &  \phi^0_d(+,-)\end{array}\right);\quad
\chi_L\equiv \left(\begin{array}{c} \chi^0_L(-,+) \\
   \chi^-_L(-,+)\end{array}\right); \quad
\chi_R\equiv \left(\begin{array}{c} \chi^0_R(+,+) \\
   \chi^-_R(+,-)\end{array}\right) .
 \end{eqnarray}

 To understand the origin of neutrino mass operators given
below, note that the gauge symmetry is broken by Higgs pair $\chi_{L,R}$
which are $SU(2)$ doublets and the usual bidoublet of the left-right model
$\phi(2,2,0)$. The left handed neutrino is part of the $SU(2)_L$ doublet
called $\psi_1$.

To see how neutrino mass arises in this theory, let us note that
there are no renormalizable operators (in the 4-D sense) that can generate
neutrino mass.
Secondly, there are three classes of  non-renormalizable operators of
higher
dimensions that remain invariant under all the symmetries of the theory,
and
which  contribute to neutrino masses.

(i) There are operators connecting
the active left handed neutrinos to themselves: i.e.
$O_1 \equiv \psi_1^TC_5 \psi_1 \phi\phi \chi_R\chi_R/M^5_*$, where
$C_5\equiv\gamma^0\gamma^2\gamma^5$, where we have omitted the family
index.
Notice that it has dimension 10 on 5D.
It generates, at the four
dimensional theory, the effective couplings
\begin{eqnarray}
{h\over (M_*R)^2}~ {(L\phi_uv_R)^2\over M_*^3  };
\end{eqnarray}
with $h$ the dimensionless coupling.
This operator induces a sufficiently
small Majorana neutrino mass,
\begin{eqnarray}
\label{numass}
m_\nu = {h~v_{wk}^2 v_R^2 \over ({M_*}R)^2~ M_*^3 } \approx ~h\cdot~ 1 ~eV
.
\end{eqnarray}
where the right hand side has been estimated using
$v_R\approx 1/R\approx 1~TeV$ and $M_*\approx 100$ TeV.
A soft hierarchy in the couplings (say $h\sim 0.01$) should provide the
right spectrum on neutrino masses.

(ii) The second class of operators connect $\nu$ to $\nu_s$ and have the
form in lowest order
$O_2 \equiv \psi^T_1\tau_2\phi \chi_R
\chi^T_R \tau_2 C_5\psi'_2/M^{7/2}_*$. This operator after
compactification has a magnitude
$\simeq \frac{v_{wk} v^2_R}{M^2_*(M_*R)^{3/2}}\simeq 10$ keV

(iii) The last class connects the left handed neutrinos that transform
under the $SU(2)_R$ group to themselves and have the form
 $O_3\equiv (\psi'_2\chi_R)^2/M^2_*$. They contribute to the
$\nu_s$-$\nu_s$ entry and have magnitude after compactification
estimated to be $\simeq \frac{v^2_R}{M^2_*R}\simeq 1-10$ GeV. The full
$6\times 6$ $\nu-\nu_s$ mass matrix
has a seesaw like form
and on diagonalization, leads to an effective mass for the light neutrino
in
the range of $0.1$ eV or so. 

This model has no twoer of bulk neutrinos with mass gap of milli eV
type; secondly, there is no need to invoke global $B-L$ symmetry to
prevent undesirable terms. This argument also carries over to the 6-D
extension of this model, which one may want for the purpose of suppressing
proton decay.

\section{Conclusions and outlook}
At the moment, neutrino oscillation experiments have provided the first
evidence for new physics beyond the standard model. The field of neutrino
physics therefore has become central to the study of new physics at
the TeV scale and beyond. The
other area which most theorists believe will be the next one to emerge
from experiments is supersymmetry. We have therefore assumed supersymmetry
in most of our discussions, although in the last section, we consider low
scale extra dimensional models without supersymmetry. 

What have we learned so far ? One thing that seems very clear is that
there is probably a set of three right handed neutrinos which restore
quark lepton symmetry to physics; secondly there must be a local $B-L$
symmetry
at some high scale beyond the standard model that is responsible
for the RH neutrinos being so far below the Planck scale. While there are
very appealing arguments that the scale of
$B-L$ symmetry is close to $10^{14}$-$10^{16}$ GeV's, in models with extra
dimensions, you cannot rule out the possibility that it is around a few
TeVs. Third thing that one may suspect is that the right handed neutrino 
spectrum may be split into a heavier one and two others which are nearby.
If this suspicion is confirmed, that would point towards an $SU(2)_H$
horizontal symmetry or perhaps even an $SU(3)_H$ symmetry which breaks
into an $SU(2)_H$ symmetry (although simple anomaly considerations prefer
the first alternative).

The correct theory should explain:

(i) Why both the solar and atmospheric mixing angles are maximal ?

(ii) Why the $\Delta m^2_{\odot} \ll \Delta m^2_A$  and what is
responsible for the smallness of $U_{e3}$ ? While in the inverted
hierarchy models, the smallness of $U_{e3}$ is natural, in general it is
not.

(iii) What is the nature of CP phases in the lepton sector and what is
their relation to the CP phases possibly responsible for baryogenesis via
leptogenesis ?

(iv) What is the complete mass psectrum for neutrinos ?

These and other questions is likely to prove to be very exciting
challenges to both theory and experiment in neutrino physics for the next
two decades.

\bigskip
\bigskip

 \noindent{\bf Acknowledgement}
This work is supported by the National Science Foundation grant number
PHY-0099544. I would like to thank Alexei Smirnov for the invitation and
kind hospitality at ICTP during the school. I like to thank A. Smirnov,
G. Senjanovi\'c, A. Perez-Lorenzana, B. Kayser, G. Fuller and E. Akhmedov
for discussions during the school. Finally, this set of lectures is
not meant to be an exhaustive review but rather a sketchy overview of
some of the important topics which is my only defense for the rather
incomplete selection of topics and cited papers.

\end{document}